\documentclass{SciPost}

\hypersetup{
    colorlinks,
    linkcolor={red!50!black},
    citecolor={blue!50!black},
    urlcolor={blue!80!black}
}

\usepackage[bitstream-charter]{mathdesign}
\DeclareSymbolFont{usualmathcal}{OMS}{cmsy}{m}{n}
\DeclareSymbolFontAlphabet{\mathcal}{usualmathcal}

\usepackage{bm}
\usepackage{dsfont}
\usepackage{booktabs}
\usepackage[section]{placeins}
\usepackage{comment}

\newcommand{\gt}{\tilde\gamma}
\newcommand{\Gt}{\tilde\Gamma}
\newcommand{\xit}{\tilde\xi}
\newcommand{\id}{\mathds{1}}
\newcommand{\rst}{\varrho_{\rm st}}
\newcommand{\rsc}{\nu_{\rm sc}}
\newcommand{\Fst}{F_{\rm st}}
\newcommand{\Hst}{H_{\rm st}}

\newcommand{\tr}{\operatorname{tr}}
\newcommand{\dd}{\mathrm{d}}
\newcommand{\artanh}{\operatorname{artanh}}
\newcommand{\Kel}{\mathsf{K}}
\newcommand{\Eel}{\mathsf{E}}
\newcommand{\Lz}{\mathcal L_0}
\newcommand{\Lo}{\mathcal L_1}
\newcommand{\av}[1]{\big\langle #1 \big\rangle}
\newcommand{\avE}[1]{\big\langle #1 \big\rangle_{\!\nu}}
\newcommand{\TT}{\mathcal T}

\def\Xint#1{\mathchoice
  {\XXint\displaystyle\textstyle{#1}}%
  {\XXint\textstyle\scriptstyle{#1}}%
  {\XXint\scriptstyle\scriptscriptstyle{#1}}%
  {\XXint\scriptscriptstyle\scriptscriptstyle{#1}}%
  \!\int}
\def\XXint#1#2#3{{\setbox0=\hbox{$#1{#2#3}{\int}$}\vcenter{\hbox{$#2#3$}}\kern-.5\wd0}}
\def\dashint{\Xint-}

\fancypagestyle{SPstyle}{
\fancyhf{}
\lhead{\colorbox{scipostblue}{\bf \color{white} ~SciPost Physics }}
\rhead{{\bf \color{scipostdeepblue} ~Submission }}

\fancyfoot[C]{\textbf{\thepage}}
}

\begin{document}

\pagestyle{SPstyle}

\begin{center}{\Large \textbf{\color{scipostdeepblue}{
Spectrum of the steady-state distribution function of disordered Lindbladian fermions at large channel number
}}}\end{center}

\begin{center}\textbf{
Maksim Parfenov\textsuperscript{1,2$\star$} and Igor Burmistrov\textsuperscript{1,2}
}\end{center}

\begin{center}
{\bf 1} L. D. Landau Institute for Theoretical Physics, Semenova 1-a, 142432, Chernogolovka, Russia
\\[\baselineskip]
{\bf 2} Laboratory for Condensed Matter Physics, HSE University, 101000 Moscow, Russia
\\[\baselineskip]
$\star$ \href{mailto:maksimqwery@gmail.com}{\small maksimqwery@gmail.com}
\end{center}

\section*{\color{scipostdeepblue}{Abstract}}
\textbf{\boldmath{%
Free fermions coupled to particle gain and loss relax to a Gaussian steady state, in which every
natural orbital has its own occupation. We find the distribution of these occupations $\varrho_{\rm st}$ for a random
Hamiltonian coupled to many random gain and loss channels. Each orbital is filled and emptied at its
own rates, and the Hamiltonian, moving particles across the levels during their lifetime, averages
the differences between the orbitals. We show that this averaging is incomplete: the spread of the
occupations decreases only as the inverse square root of the number of channels per level. At large
system size the distribution depends on the grain Hamiltonian only through its density of states $\nu(E)$. The spread is bounded by the baths alone: even
Hamiltonians with unbounded, heavy-tailed spectra cannot widen it beyond this bound. The spread suppresses the particle-number fluctuations
below their binomial value.}}

\vspace{\baselineskip}

\vspace{10pt}
\noindent\rule{\textwidth}{1pt}
\tableofcontents
\noindent\rule{\textwidth}{1pt}
\vspace{10pt}

\section{Introduction}\label{sec:intro}
Anderson localisation in open quantum systems has recently attracted much attention. The
coupling to an environment makes the dynamics non-unitary. It is generated by a non-Hermitian
Hamiltonian~\cite{HatanoNelson1996,KawabataRyu2021,Luo2021} or, for the full density matrix, by a
Lindbladian~\cite{Yusipov2017}. A central question is then which quantities replace the spectral
characteristics that detect localisation in closed systems.

In a closed system the Anderson transition is conveniently detected by the level statistics of the
Hamiltonian. The reference point is the zero-dimensional limit, the random-matrix theory. It follows
from the zero mode of the nonlinear $\sigma$-model~\cite{Efetov1983,Efetov1997} and gives the
Wigner--Dyson statistics of extended states, while deviations from it signal the approach to
localisation~\cite{Shklovskii1993,EversMirlin2008}.

For open systems of free fermions a natural analogue of the Hamiltonian exists. If the Lindbladian is
quadratic, i.e.\ the jump operators are linear in the fermion operators, and its steady state is
unique, this state is Gaussian~\cite{Prosen2008}, $\hat\rho\propto\exp(-\hat c^\dagger\Hst\hat c)$. It is fixed by the
single-particle matrix $\Hst$, the modular Hamiltonian. Its eigenvalues $\beta_j$, the modular
energies, and its eigenvectors, the natural orbitals, play the role of the energies and eigenstates of
a closed system. As in the closed case, the first step is the zero-dimensional limit, i.e.\ the
statistics of $\Hst$ for a random Lindbladian. Spectra and steady states of random Lindbladians have
been recently studied in Refs.~\cite{Denisov2019,Can2019,Sa2020,Costa2023}. Thompson and
Kamenev~\cite{Thompson2023} introduced a zero-dimensional model of this kind, a random Hamiltonian
coupled to random gain and loss channels, and formulated its Keldysh field theory.

However, the distribution of the modular energies in this model has not been found analytically.
At a large number of channels it was heuristically described as a semicircle of width $\propto1/m$, where
$m=M/N$ is the number of channels per level~\cite{Thompson2023}. In this paper we compute the average
density of the modular energies
\begin{equation}\label{eq:rst}
\rst(\beta)=\left\langle\frac1N\sum_{j=1}^N\delta(\beta-\beta_j)\right\rangle
\end{equation}
at $m\gg1$.

The form of $\rst$ is set by a competition between the bath noise and the Hamiltonian. In the basis
of the natural orbitals $\hat d_j$ the steady state factorises,
$\hat\rho\propto\exp(-\sum_j\beta_j\hat d_j^\dagger\hat d_j)$, and $\beta_j=\ln[(1-n_j)/n_j]$ is fixed by the
occupation $n_j$ of the $j$-th orbital. Unlike in the Gibbs state, where $\Hst=(H_0-E_F)/T$, the
$\beta_j$ are not energies divided by a temperature: the natural orbitals are not eigenstates of the
Hamiltonian, and no energy enters $\beta_j$. They are set by a balance of rates. An orbital that is filled by the gain at the rate
$\tilde\gamma_j$ and emptied by the loss at the rate $\gamma_j$ has $\tilde\gamma_j(1-n_j)=\gamma_jn_j$,
i.e.\ $\beta_j=\ln(\gamma_j/\tilde\gamma_j)$. If every orbital saw the mean rates $\gamma$ and
$\tilde\gamma$ of the loss and gain channels, all $\beta_j$ would coincide,
$\beta_j=\beta_0=\ln(\gamma/\tilde\gamma)$, and $\rst$ would be a single delta peak. With $M=mN$
random channels the loss and gain rate matrices $\Gamma$ and $\Gt$ fluctuate: their eigenvalues spread by
a relative amount of order $m^{-1/2}$, and the natural orbitals adjust to these fluctuations. The
Hamiltonian, which moves a particle across the levels during its lifetime, averages these
fluctuations (Sec.~\ref{sec:resum}). The distribution \eqref{eq:rst} measures how much of the bath
noise survives this averaging. We find that its width scales as $m^{-1/2}$ rather than $1/m$, and
that its shape depends on the Hamiltonian only through its density of states.

The outline of the paper is as follows. In Sec.~\ref{sec:model} we formulate the model. In
Sec.~\ref{sec:resum} we reduce $\rst$ at $m\gg1$ to the spectral density of a single random matrix.
Its width, shape and edge are found in Secs.~\ref{sec:width}, \ref{sec:wigner} and \ref{sec:edge}. In
Sec.~\ref{sec:confine} we bound the modular energies and consider L\'evy Hamiltonians. Observable
consequences are discussed in Sec.~\ref{sec:fluct}, and the relation to earlier work in
Sec.~\ref{sec:earlier}. We conclude in Sec.~\ref{sec:conclusion}. Technical details are given in
Appendices.

\section{Model and the bath fluctuations}\label{sec:model}
\begin{figure}[t]
\centering
\includegraphics{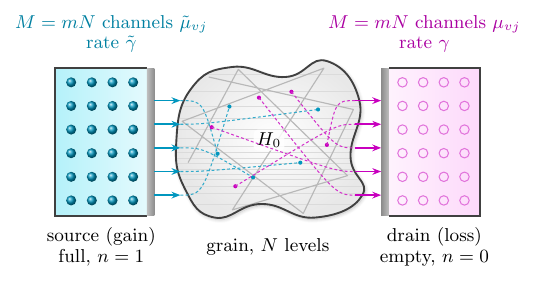}
\caption{$N$ fermionic levels between a full source and an empty drain, each coupled through
$M=mN$ channels. Every channel couples to a random superposition of the levels. The gain channels
$\tilde\mu_{vj}$ fill the levels at the rate $\gt$, the loss channels $\mu_{vj}$ empty them at the
rate $\gamma$.}
\label{fig:setup}
\end{figure}
We consider $N$ fermionic modes $\hat c_j$ evolving under a Hamiltonian $\hat{H}_0=\hat c^\dagger H_0\hat c$ and coupled to
a full source and an empty drain (Fig.~\ref{fig:setup}), each through $M=mN$ channels. The density matrix of such a system obeys the Gorini--Kossakowski--Sudarshan--Lindblad (GKSL) master equation 
\cite{Lindblad1976,GKS1976}:
\begin{equation}
 \partial_t \rho = -i[\hat{H}_0,\rho] +\sum_{a,v}\Bigl (
 \hat{L}_{av}\rho \hat{L}_{av}^\dag-\frac{1}{2}\Bigl\{\hat{L}_{av}^\dag\hat{L}_{av},\rho\Bigr\}
\Bigr ) .
\label{eq:GKSL}
\end{equation}
Here the jump
operators are given by
\begin{equation}
\hat L_{1v}=\sum_j\mu_{vj}\,\hat c_j,\qquad \hat L_{2v}=\sum_j\tilde\mu_{vj}\,\hat c_j^\dagger,
\qquad v=1,\dots,M=mN ,
\end{equation}
with independent complex Gaussian entries, $\langle|\mu_{vj}|^2\rangle=\gamma/M$ and
$\langle|\tilde\mu_{vj}|^2\rangle=\gt/M$: as in the random-matrix theory of chaotic quantum
dots~\cite{Mahaux1969,Beenakker1997,Alhassid2000}, a channel couples to a random superposition of the
modes~\cite{Berry1977}. The loss and gain dissipative fields can be conveniently combined into the $N\times N$ Wishart matrices \cite{Thompson2023}
\begin{equation}\label{eq:GG}
\Gamma=\mu^\dagger\mu,\qquad \Gt=(\tilde\mu^\dagger\tilde\mu)^{\mathsf T} .
\end{equation}
We also introduce the $N\times N$ matrices
\begin{equation}\label{eq:QD}
Q=\tfrac12(\Gamma+\Gt),\qquad D=\Gamma-\Gt,\qquad H=H_0-iQ .
\end{equation}
The total rate of gain and loss, $s=\gamma+\gt$, is the relaxation rate of the occupations, while single-particle amplitudes decay at the rate $s/2$.
Since $\tilde\mu$ is a complex Ginibre matrix, the transpose in Eq. \eqref{eq:GG} does not affect the statistical properties 
of $\Gt$
and, therefore, plays no role below.

It is convenient to write the matrices $\Gamma$ and $\Gt$ as their means times relative fluctuations:
\begin{equation}\label{eq:split}
\Gamma=\gamma(\id+\xi),\qquad \Gt=\gt(\id+\xit).
\end{equation}
To set the stage we discuss the properties of the random fluctuation matrices $\xi$, $\xit$.
\begin{enumerate}
\item[(a)] \emph{The mean is exactly a multiple of the identity}. Since
$\langle\mu^*_{vj}\mu_{wk}\rangle=(\gamma/M)\delta_{vw}\delta_{jk}$, we have
$\langle\Gamma\rangle=\gamma\id$ at any finite $N,M$, so $\langle\xi\rangle=\langle\xit\rangle=0$.
\item[(b)] \emph{Their correlators are exactly known.} Wick's theorem for the circular Gaussian
weight gives (Appendix~\ref{app:wishart})
\begin{equation}\label{eq:2pt}
\av{\xi_{jk}\xi_{lm}}=\frac1M\delta_{jm}\delta_{kl},\qquad
\av{\xi_{jk}\xit_{lm}}=0,
\end{equation}
while all higher cumulants are suppressed by further powers of $1/m$. The relevant parameter is
$1/m=N/M$, not $1/M$: each extra cumulant order carries one more factor $1/M$ but also one more free
matrix index to be summed over, hence a factor $N$. The counting is done explicitly in Appendix~\ref{app:wishart}.
\item[(c)] \emph{$\xi$ and $\xit$ are unitarily invariant.} Let $U$ be a fixed unitary  matrix,
$U^{\dagger} U = \id$, independent of $\mu$, and $\xi'=U^{\dagger} \xi U$. Then, we find
\begin{equation}
\big\langle  \xi'_{jk} \xi'_{lm}\big\rangle
=\sum_{abcd}U^*_{aj}U_{bk}U^*_{cl}U_{dm}\big\langle \xi_{ab}\xi_{cd}\big\rangle
=\frac1M\Big(\sum_aU^*_{aj}U_{am}\Big)\Big(\sum_bU^*_{bl}U_{bk}\Big)=\frac1M\,\delta_{jm}\delta_{kl},
\label{eq:item:c}
\end{equation}
We note that the tensor structure $\delta_{jm}\delta_{kl}$ reproduces
itself under any $\mu$-independent unitary transformation. This property can also be used for higher-order correlators.

\end{enumerate}

The Hamiltonian $H_0$ is an arbitrary Hermitian random matrix, independent of the couplings $\mu,\tilde\mu$,
whose level density converges for $N\to\infty$ to a density of states $\nu(E)$, normalised to one, $\int dE \nu(E)=1$. 
Our main example is the Gaussian unitary ensemble (GUE), $\langle|(H_0)_{jk}|^2\rangle=1/N$, whose density of states is the
Wigner semicircle of radius $2$,
\begin{equation}\label{eq:nusc}
\nu(E)=\nu_{\rm sc}(E)=\frac{\sqrt{4-E^2}}{2\pi},\qquad W=4 .
\end{equation}
Here $W$ denotes the width of the spectrum. In Secs.~\ref{sec:universal} and \ref{sec:confine} we consider also the Gaussian orthogonal ensemble (GOE), Hamiltonians with
uncorrelated or with rigid levels, as well as Hamiltonians drawn from L\'evy matrix ensembles with unbounded, heavy-tailed spectra.

Since the jump operators are linear in $\hat c$, Eq.~\eqref{eq:GKSL} closes on the
one-particle distribution matrix $F_{jk}=\langle[\hat c_j,\hat c_k^\dagger]\rangle=\delta_{jk}-2\langle\hat
c_k^\dagger\hat c_j\rangle$, which obeys the quantum kinetic equation~\cite{Prosen2008,Thompson2023}
\begin{equation}\label{eq:kinetic0}
\partial_tF=-i\big(HF-FH^\dagger\big)+D,
\end{equation}
with $H$ and $D$ from Eq.~\eqref{eq:QD}. Its stationary solution, the steady-state distribution matrix $\Fst=\tanh(\Hst/2)$, solves the
Lyapunov-Sylvester equation~\cite{Prosen2008}
\begin{equation}\label{eq:lyap}
H\Fst-\Fst H^\dagger=-iD
\qquad\Longleftrightarrow\qquad
 i[H_0,\Fst]+\{Q,\Fst\}=D,
\end{equation}
where $\Hst$ is the modular Hamiltonian of the steady state, $\hat\rho\propto e^{-\hat c^\dagger\Hst\hat c}$.
Its eigenvalues, the modular energies $\beta_j=2\artanh f_j$, where $f_j$ are the eigenvalues of
$\Fst$, are the subject of this paper. In the basis of the natural orbitals $\hat d_j$ the state
factorises, $\hat\rho=\prod_je^{-\beta_j\hat d_j^\dagger\hat d_j}/(1+e^{-\beta_j})$, so that
$\beta_j=\ln[(1-n_j)/n_j]$ is the log-odds of the occupation $n_j$ of the $j$-th natural orbital, and
$e^{-\beta_j}$ is its effective fugacity.

To characterise the steady state we use the average density \eqref{eq:rst} of the modular
energies. It is normalised as a probability density, and $\langle\cdots\rangle$ denotes the average
over $H_0$, $\mu$ and $\tilde\mu$.

\begin{figure}[t]
\centering
\includegraphics{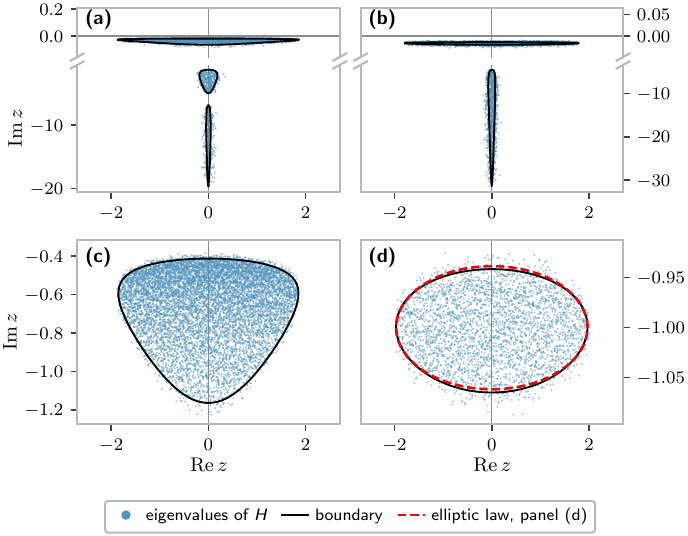}
\caption{Eigenvalues of $H=H_0-iQ$ for the GUE $H_0$ (points) and the boundary \eqref{eq:cloudbdy}
(black line). In (a) and (b) the vertical axis is broken: the upper strip shows the piece next to the real axis, which is thinner than a line on the scale of the full cloud. (a) Three pieces at $m=0.07$, $\gamma=1.7$, $\gt=0.5$ ($N=400$, $12$ realisations).
(b) Two pieces at $m=0.1$, $\gamma=\gt=3$ ($N=400$, $12$ realisations). (c) One piece at $m=1.5$,
$\gamma=1$, $\gt=0.3$ ($N=300$, $30$ realisations). (d) One piece at $m=20$, $\gamma=1.5$, $\gt=0.5$
($N=300$, $10$ realisations), together with the elliptic law \eqref{eq:ellipse} (red, dashed).}
\label{fig:cloud}
\end{figure}

\subsection{Non-Hermitian dynamics}\label{subsec:cloud}

The matrix $H=H_0-iQ$ in Eq.~\eqref{eq:lyap} is the effective single-particle Hamiltonian of the
open system. The solution of the kinetic equation \eqref{eq:kinetic0} is
\begin{equation}\label{eq:kinetic}
F(t)-\Fst=e^{-iHt}\big[F(0)-\Fst\big]e^{iH^\dagger t}, \quad \Fst=\int\limits_0^\infty\dd t\,e^{-iHt}De^{iH^\dagger t} .
\end{equation}
The steady state is thus reached if all
eigenvalues $z_j$ of $H$ lie in the lower half-plane. Then $-\mathrm{Im}\,z_j$ are the decay rates of the
single-particle amplitudes, and the modes of $F$ relax with the rates $-\mathrm{Im}\,(z_j+z_k)$. At $m=\infty$ the 
imaginary part of $H$
becomes a multiple of the identity, $Q=\frac s2\id$
(see Eq.~\eqref{eq:QD1} below) such that $z_j=E_j-is/2$, where $E_j$ are eigenenergies of $H_0$. Therefore, for $m=\infty$ all amplitudes decay at the same rate $s/2$, and $F$ relaxes at the rate $s$. At finite $m$ the
fluctuations of $Q$ spread the eigenvalues of $H$ into a cloud in the complex plane
(see Fig.~\ref{fig:cloud}). For $N\to\infty$ and $H_0\in$ GUE  the boundary of the cloud is the curve which can be found from the following equations (see Appendix~\ref{app:sigma}, where they follow from the saddle point of the disorder-averaged Keldysh action):
\begin{equation}\label{eq:cloudbdy}
\frac1{|G|^2}=1+\sum_{\sigma=\pm}\frac{\gamma_\sigma^2}{4m\,|\alpha_\sigma|^2},\qquad
\alpha_\sigma=1+\frac{i\gamma_\sigma}{2m}\,G,
\qquad \gamma_+=\gamma, \quad \gamma_-=\gt .
\end{equation}
Here $G(z)=\av{\frac1N\tr(z-H)^{-1}}$ is the averaged resolvent. We note that 
$G(z)\simeq1/z$ at $|z|\to\infty$.  
For $N\to \infty$ the averaged resolvent satisfies the self-consistent equation
\begin{equation}\label{eq:Gcloud}
\frac1G=z-G+\sum_{\sigma=\pm}\frac{i\gamma_\sigma}{2\alpha_\sigma}.
\end{equation}

In the regime $m\gg1$ studied in this paper, Eq. \eqref{eq:Gcloud} reduces to the algebraic equation $1/G=z+is/2-(1-v)G$. Then the cloud becomes the ellipse determined by  
\begin{equation}\label{eq:ellipse}
\frac{(\mathrm{Re}\,z)^2}{4}+\frac{(\mathrm{Im}\,z+s/2)^2}{4v^2}=\frac{1}{1+v},\qquad
v=\frac{\gamma^2+\gt^2}{4m}=N\av{|(Q_1)_{jk}|^2}.
\end{equation}
We note that Eq. \eqref{eq:ellipse} is nothing but
the elliptic law~\cite{Sommers1988,Fyodorov1997} for $H_0-iQ_1$, where $Q_1$ is the fluctuating part of $Q$, see Eqs.~\eqref{eq:2pt} and \eqref{eq:QD1} below. The height of the cloud is of order $1/m$, although $Q_1$ is
of order $m^{-1/2}$. The Hamiltonian $H_0$ mixes the modes faster than the baths attenuate them, and the amplitude decay rates are
averaged to $s/2$ (the so-called motional narrowing). The modular energies are not averaged in this way: their
spread is of order $m^{-1/2}$ (see Sec.~\ref{sec:width}). At $m<1/2$, the matrices $\Gamma$ and $\Gt$
have rank $M<N/2$, and $Q$ has rank $2M<N$. If the baths are coupled strongly enough, the cloud splits into two parts as shown in Fig.~\ref{fig:cloud}(b). $2M$ modes are damped
directly, with $-\mathrm{Im}\,z\simeq\gamma/2m$ and $\gt/2m$, while the remaining $N-2M$ decay only
through their mixing with them. Therefore, these modes stay close to the real axis,
$-\mathrm{Im}\,z\simeq2m^2(1/\gamma+1/\gt)/(1-2m)$ (an effect similar to the quantum Zeno effect).

If $\gamma/\gt$ is large enough, the directly damped modes split further into $M$ modes
damped mainly by the loss and $M$ modes damped mainly by the gain, and the cloud consists of three
pieces [Fig.~\ref{fig:cloud}(a), $\gamma/\gt=3.4$].

For a single bath these features are known. At $\gt=0$ Eq.~\eqref{eq:cloudbdy} gives the
resonance cloud of a chaotic cavity with $M$ open channels~\cite{Haake1992,Janik1997}, and its
splitting is the resonance trapping transition~\cite{Sokolov1989}. 
The boundary for two baths
with different rates and its derivation from the Keldysh action (Appendix~\ref{app:sigma}) are new results to our knowledge. For the
rest of the paper we need only the large-$m$ result: all amplitudes decay with the same rate $s/2$, whereas
the modular energies keep fluctuations of order $m^{-1/2}$.

\section{Solution of the Lyapunov-Sylvester equation}\label{sec:resum}
Let us now return to the calculation of $\rst$. To this end, we need to solve the Lyapunov-Sylvester equation \eqref{eq:lyap}. Let us split the matrices $Q$ and $D$ into their mean values and fluctuations,
\begin{equation}\label{eq:QD1}
Q=\frac s2\id+Q_1,\quad Q_1=\tfrac12(\gamma\xi+\gt\xit),\qquad
D=(\gamma-\gt)\id+D_1,\quad D_1=\gamma\xi-\gt\xit,
\end{equation}
where $s=\gamma+\gt$. Eq.~\eqref{eq:lyap} becomes, exactly,
\begin{equation}\label{eq:exact}
\big(\Lz+\Lo\big)\Fst=D,\qquad
\Lz A=i[H_0,A]+sA,\qquad \Lo A=\{Q_1,A\}.
\end{equation}
We solve it by perturbation theory in powers of $m^{-1/2}$, $\Fst=F_0+F_1+\dots$, where $F_n$ is of
order $m^{-n/2}$.

The identity matrix is an eigenvector of $[H_0,\cdot]$ with eigenvalue zero for every $H_0$, so
$\Lz\id=s\id$. Because the zeroth-order source is exactly $D_0=(\gamma-\gt)\id$ by Eq.~\eqref{eq:QD1}, we find
\begin{equation}\label{eq:F0}
F_0=\frac{\gamma-\gt}{s}\,\id,\qquad
\beta_0=2\artanh\frac{\gamma-\gt}{\gamma+\gt}=\ln\frac{\gamma}{\gt},\qquad
n_0=\frac{\gt}{s}.
\end{equation}
At $m=\infty$ every natural orbital has the same modular energy $\beta_0$. The state
$e^{-\beta_0\hat N}$ is then the state in which every orbital has the same occupation $n_0=\gt/s$, set
by the balance of the rates, $e^{-\beta_0}=\gt/\gamma$, and it keeps no memory of $H_0$. As we demonstrate below, the nontrivial form of $\rst$ describes the spreading of this $N$-fold degenerate point at finite $m$.

Sorting Eq. \eqref{eq:exact} by powers of $m^{-1/2}$ gives
$\Lz F_1=D_1-\{Q_1,F_0\}=D_1-2F_0Q_1$, where we used that $F_0$ is proportional to the identity matrix. Using
relations \eqref{eq:QD1}, we find that the loss and gain channels enter with the same prefactor and opposite
signs:
\begin{equation}\label{eq:F1}
D_1-2F_0Q_1=\frac{2\gamma\gt}{s}\,X,\qquad X=\xi-\xit,\qquad
F_1=\frac{2\gamma\gt}{s}\,\Lz^{-1}X .
\end{equation}
If $\xi=\xit$, the scalar $F_0$ solves Eq. \eqref{eq:exact} exactly, and $\rst=\delta(\beta-\beta_0)$ for
every $m$. The width of $\rst$ is generated solely by the mismatch between the two baths, and it is a
purely non-equilibrium effect.

Since $F_0$ is proportional to the identity matrix, $H_{\rm st}=2\artanh F_{\rm st}$ may be linearised with the ordinary Taylor series:
\begin{equation}\label{eq:master}
\Hst=\beta_0\,\id+Y+O(1/m),\qquad
Y=s\,\Lz^{-1}X.
\end{equation}
The inverse of $\Lz$ is a Laplace transform. For a matrix evolved with the Hamiltonian,
$A(\tau)=e^{-iH_0\tau}Ae^{iH_0\tau}$, one has $i[H_0,A(\tau)]=-\partial_\tau A(\tau)$, so that, after
an integration by parts,
\begin{equation}
\Lz\int_0^\infty\dd\tau\,e^{-s\tau}A(\tau)=\int_0^\infty\dd\tau\,e^{-s\tau}(s-\partial_\tau)A(\tau)=A .
\end{equation}
Hence
\begin{equation}\label{eq:Ytime}
Y=s\int_0^\infty\dd\tau\,e^{-s\tau}\,X(\tau),\qquad X(\tau)=e^{-iH_0\tau}\,X\,e^{iH_0\tau}.
\end{equation}
The spread of the modular energies is the bath noise $X$, evolved by the Hamiltonian and averaged
over the relaxation time $1/s$ of the occupations.
The centre of the distribution depends only on the ratio $\gamma/\gt$ whereas, at fixed $m$ and $\nu$, the fluctuations depend on the bath rates only through $s$.

The spectrum of $Y$ gives the leading large-$m$ approximation to $\rst$, written below as a function of the shifted modular energy
$b=\beta-\beta_0$. At
$m=\infty$ all $N$ eigenvalues of $\Hst$ coincide, $\beta_j=\beta_0$ for every $j$, cf.
Eq.~\eqref{eq:F0}, i.e., the distribution is a single delta peak $\rst=\delta(\beta-\beta_0)$. The matrix $Y$, of typical size
$m^{-1/2}$, is the perturbation that lifts this exactly $N$-fold degenerate level. 

\begin{figure}[t]
\centering
\includegraphics{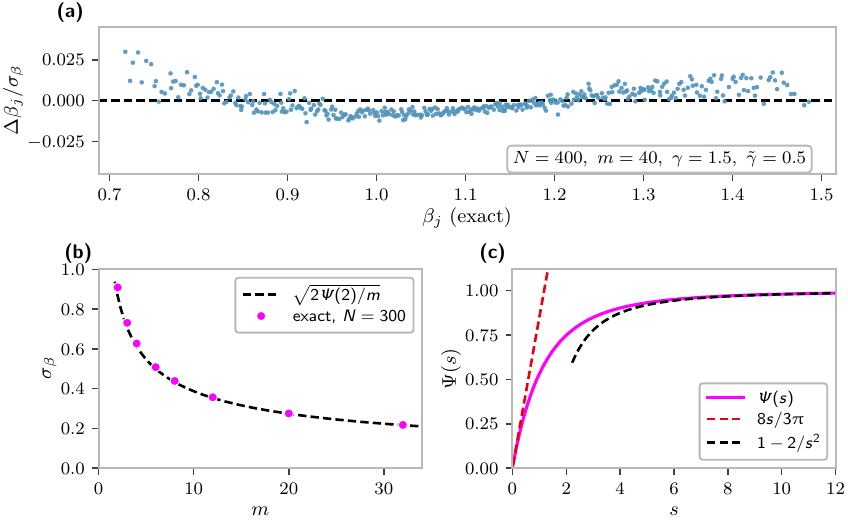}
\caption{Validity of the leading order and the width. (a) Difference $\Delta\beta_j$ between the sorted
eigenvalues of the exact $\Hst$ and of $\beta_0+Y$, Eq.~\eqref{eq:master}, for one realisation ($N=400$,
$m=40$, $\gamma=1.5$, $\gt=0.5$), in units of the width $\sigma_\beta=\sqrt{2\Psi(2)/m}=0.1935$ of
Eqs.~\eqref{eq:width} and \eqref{eq:psi}. It does not exceed $0.03\,\sigma_\beta$ for every eigenvalue.
(b) Width of $\rst$ against $m$ at $\gamma=1.5$, $\gt=0.5$. Points: exact Lyapunov solutions ($N=300$,
$4$ realisations). Dashed line: $\sqrt{2\Psi(2)/m}$. (c) The closed form \eqref{eq:psi} for $\Psi(s)$
(solid) with its small-$s$ (red, dashed) and large-$s$ (black, dashed) asymptotes \eqref{eq:asym}.}
\label{fig:width}
\end{figure}

Figure~\ref{fig:width}(a) confirms this picture against the exact numerical solution of the Lyapunov-Sylvester equation~\eqref{eq:lyap}.
The rest of the paper is devoted to the study of that single matrix $Y$: its second moment
(Sec.~\ref{sec:width}), the shape of its spectral density (Sec.~\ref{sec:wigner}) and the edge of the spectral density
(Sec.~\ref{sec:edge}).

\section{The width of the distribution}\label{sec:width}
\subsection{Laplace representation}\label{subsec:width}
We first compute the second moment of $\rst$, i.e.\ of the spectrum of $Y$, Eq.~\eqref{eq:Ytime}.
Since $\frac1N\tr Y=\frac1N\tr X$ has zero mean, $\av\beta=\beta_0$ at this order, and
\begin{equation}
\sigma_\beta^2=\Big\langle\frac1N\tr\Hst^2\Big\rangle-\Big\langle\frac1N\tr\Hst\Big\rangle^2
=\Big\langle\frac1N\tr Y^2\Big\rangle
=s^2\!\int_0^\infty\!\!\dd\tau\!\int_0^\infty\!\!\dd\tau'\,e^{-s(\tau+\tau')}\,
\Big\langle\frac1N\tr X(\tau)X(\tau')\Big\rangle .
\end{equation}
The averages over $H_0$ and over the baths are independent, so we perform averaging over the baths first. By using Eq.
\eqref{eq:2pt} and the relation $\av{\xi\,\xit}=0$, we find $\av{\tr AXBX}_{\mu,\tilde\mu}=\frac2M\tr A\,\tr B$ for any matrices
$A,B$ independent of the baths. Hence, with $A,B$ being the evolution operators, we obtain
\begin{equation}
\Big\langle\frac1N\tr X(\tau)X(\tau')\Big\rangle_{\mu,\tilde\mu}=\frac2m\,\big|\mathcal R(\tau-\tau')\big|^2,
\qquad \mathcal R(\tau)=\frac1N\tr e^{-iH_0\tau}.
\end{equation}
$\mathcal R(\tau)$ is the return amplitude of $H_0$, the average overlap of a random state with its own time
evolution. The integrand depends only on $\tau-\tau'$ and is even in it. Then we can simplify this relation by using the identity, 
$s^2\!\int\!\!\int\dd\tau\,\dd\tau'e^{-s(\tau+\tau')}f(\tau-\tau')=s\!\int_0^\infty\dd\tau\,e^{-s\tau}f(\tau)$.
For
$N\gg1$ the level density converges to $\nu(E)$, and $\mathcal R(\tau)$ to its Fourier
transform $\hat\nu(\tau)=\int\dd E\,\nu(E)\,e^{-iE\tau}$. Altogether,
\begin{equation}\label{eq:width}
\sigma_\beta^2=\frac2m\,\Psi(s),\qquad
\Psi(s)=s\int_0^\infty\dd\tau\,e^{-s\tau}\,\big|\hat\nu(\tau)\big|^2 .
\end{equation}
The variance is set by the Laplace transform, at the relaxation rate $s$, of the squared return amplitude. The
baths kick the modes, $H_0$ dephases the kick, and only the part that has not dephased within the
relaxation time $1/s$ survives. Writing $|\hat\nu|^2$ as a double integral over two levels and performing the $\tau$ integral
gives the equivalent form
\begin{equation}\label{eq:psidef}
\Psi(s)=\iint\dd E\,\dd E'\,\nu(E)\,\nu(E')\,\frac{s^2}{(E-E')^2+s^2}
=\Big\langle\frac{s^2}{(E-E')^2+s^2}\Big\rangle_{\!\nu},
\end{equation}
where $\langle\cdots\rangle_\nu$ averages over two levels drawn independently from $\nu$.

It is worthwhile to highlight the universality of this answer. The resulting width depends on $H_0$ only through its density of states. The
eigenvectors of $H_0$ never entered: the bath average $\av{\tr AXBX}$ is the same for any $A,B$
independent of the baths. Level correlations enter only at finite $N$, through the average of
$|\mathcal R(\tau)|^2$, i.e.\ the spectral form factor.

\subsection{The GUE: closed form}\label{sec:gueclosed}

For the semicircle \eqref{eq:nusc} the return amplitude is
$\hat\nu_{\rm sc}(\tau)=J_1(2\tau)/\tau$, and
\begin{equation}\label{eq:bessel}
\Psi(s)=s\int_0^\infty\dd\tau\,e^{-s\tau}\,\frac{J_1(2\tau)^2}{\tau^2}.
\end{equation}
This Laplace transform evaluates to the complete elliptic integrals of the
first and second kind,
$\Kel(p)=\int_0^{\pi/2}\dd\theta\,(1-p\sin^2\theta)^{-1/2}$ and
$\Eel(p)=\int_0^{\pi/2}\dd\theta\,(1-p\sin^2\theta)^{1/2}$ with parameter $p=-16/s^2$:
\begin{equation}\label{eq:psi}
\Psi(s)=\frac{s^2}{24\pi}\Big[(s^2+16)\,\Kel\!\big(-\tfrac{16}{s^2}\big)
-(s^2-16)\,\Eel\!\big(-\tfrac{16}{s^2}\big)-12\pi\Big] .
\end{equation}
The comparison of $\sigma_\beta$ with the width of exact numerical solutions is shown in
Fig.~\ref{fig:width}(b). The asymptotics are
\begin{equation}\label{eq:asym}
\Psi(s)\underset{s\gg4}{=}1-\frac2{s^2}+\frac{10}{s^4}-\frac{70}{s^6}+\dots,\qquad
\Psi(s)\underset{s\ll1}{=}\frac{8s}{3\pi}-\frac{s^2}{2}+\frac{s^3}{4\pi}\Big[\ln\frac{16}{s}-\frac12\Big]+O(s^4).
\end{equation}
Both follow directly from Eq. \eqref{eq:width}. For $s\gg4$ the weight $s\,e^{-s\tau}$ probes only short
times, where $|\hat\nu_{\rm sc}(\tau)|^2=1-\tau^2+\dots$ is set by the moments of $\rsc$. For
$s\ll1$ the weight is nearly flat, and the leading term is
$s\int_0^\infty\dd\tau\,|\hat\nu_{\rm sc}|^2=\pi s\int\dd E\,\rsc^2=8s/3\pi$, the fraction of
level pairs closer than $s$. The function $\Psi(s)$ and its two asymptotes are shown in
Fig.~\ref{fig:width}(c).

\section{The vector Dyson equation and the shape of \texorpdfstring{$\varrho_{\rm st}$}{rho\_st}}\label{sec:wigner}
The width fixes only the second moment. For the shape we first show that $Y$ is a Wigner-type
matrix whose statistics is fixed, once and for all, by one function $S(E,E')$ of the pair of $H_0$
levels it connects (Sec.~\ref{subsec:profile}). We then derive a deterministic equation for its
spectral density (Sec.~\ref{sec:dyson}), check that it depends on $H_0$ only through $\nu$
(Sec.~\ref{sec:universal}), and solve it for the GUE in the two limits $s\gg W$ and $s\ll W$
(Secs.~\ref{sec:larges} and \ref{sec:smalls}).

\subsection{The $Y$ correlator}\label{subsec:profile}
Let $H_0|j\rangle=E_j|j\rangle$, and denote by a prime a matrix written in this eigenbasis,
$X'_{jk}=\langle j|X|k\rangle$. Note that $X'_{jk}=\xi'_{jk}-\xit'_{jk}$ are the bath fluctuations
in the eigenbasis of $H_0$, not the entries $X_{jk}$ in the original basis of the modes
$\hat c_j$ that appear in Eq. \eqref{eq:2pt}. In this basis the evolution in Eq. \eqref{eq:Ytime} is
diagonal, $\langle j|X(\tau)|k\rangle=e^{-i\omega_{jk}\tau}X'_{jk}$ with $\omega_{jk}= E_j-E_k$,
and the time integral can be done entry by entry:
\begin{equation}\label{eq:rtime}
\langle j|Y|k\rangle=Y'_{jk}=r_{jk}\,X'_{jk},\qquad
r_{jk}=s\int_0^\infty\!\dd\tau\;e^{-s\tau}\,e^{-i\omega_{jk}\tau}=\frac{s}{i\omega_{jk}+s}.
\end{equation}
In the eigenbasis of $H_0$ the matrix $Y$ is thus the Hadamard product $r\circ X'$, and
every matrix operation below (products, resolvents, traces) is performed on these matrix elements.

Because $\Gamma$ and $\Gt$ are Hermitian, so is
$X=\xi-\xit$, whence $X'_{kj}=\overline{X'_{jk}}$, where an overbar denotes complex conjugation
throughout. Since $\bar r_{jk}=r_{kj}$, the matrix $Y$ is Hermitian as it must be, and since
$r_{jj}=1$, the diagonal of $Y$ is that of $X$. The baths therefore do not shift the centre of $\rst$ at this
order. The variance of the off-diagonal entries follows from
Eq. \eqref{eq:2pt}. The eigenbasis of $H_0$ is a function of $H_0$ alone and is therefore independent of
$\mu,\tilde\mu$, so by property \eqref{eq:item:c} the rotated fluctuations $X'$ have
exactly the same correlators as $X$. In particular
$\av{|X'_{jk}|^2}=\av{X'_{jk}X'_{kj}}=2/M$, the anomalous contraction $\av{X'_{jk}X'_{jk}}$ vanishes
for $j\ne k$, and the entries of $X'$ carry no correlation with the energies $E_j$. Therefore
\begin{equation}
N\av{|Y'_{jk}|^2}_{\mu,\tilde\mu}=\frac{2N}{M}\,|r_{jk}|^2=S(E_j,E_k), \qquad
S(E,E')=\frac{2}{m}\,\frac{s^2}{(E-E')^2+s^2}.
\label{eq:Sdef}
\end{equation}

It is useful to note the sum rule that controls the total coupling of one level to the rest of the
band,
\begin{equation}\label{eq:sumrule}
\frac1N\sum_kS(E_j,E_k)\;\longrightarrow\;u(E_j),\qquad
u(E)=\int\dd E'\,\nu(E')\,S(E,E')\le\frac2m ,
\end{equation}
with equality only for $s\to\infty$. Its average over $\nu$ is the variance $\sigma_\beta^2$, Eq.~\eqref{eq:width}, and the
fact that it is not constant across the band is what makes $\rst$ differ from a semicircle at
all. For the semicircle $u$ is elementary,
\begin{equation}\label{eq:u}
u(E)=\frac sm\Big[\mathrm{Im}\sqrt{(E+is)^2-4}\,-\,s\Big],
\end{equation}
with the branch fixed by $\sqrt{z^2-4}\to z$ at infinity.

\subsection{The vector Dyson equation}\label{sec:dyson}

A calculation of the shape of $\rst$ requires knowledge of the full resolvent of $Y$.
We hold $H_0$ fixed, so that its eigenvalues $E_j$ are parameters, and average over the baths
first. By property \eqref{eq:item:c} and Eq.~\eqref{eq:2pt}, the entries of $Y$
in the eigenbasis of $H_0$ are centred and uncorrelated, except for the Hermitian pairs
$Y'_{kj}=\bar Y'_{jk}$, with the variances \eqref{eq:Sdef}:
\begin{equation}\label{eq:Ycov}
\av{Y'_{jk}Y'_{lm}}=r_{jk}r_{lm}\av{X'_{jk}X'_{lm}}=\frac1N\,S(E_j,E_k)\,\delta_{jm}\delta_{kl}.
\end{equation}
The non-Gaussian cumulants of $X$ are suppressed by further powers of $1/m$ (Appendix~\ref{app:wishart}).
Hence, at the order at which \eqref{eq:master} holds, $Y$ is a Wigner-type matrix: its entries in
different pairs $\{j,k\}$ can be replaced by independent Gaussian variables with the variance profile $S$.

The density is obtained from the resolvent $G(\lambda)=(\lambda-Y)^{-1}$ by the standard relation
$\rst=-\frac1\pi\lim_{\eta\to0^+}\mathrm{Im}\frac1N\tr G(b+i\eta)$, so we need its diagonal
elements $g_j=\av{G_{jj}}$. Singling out the index $j$, the Schur complement formula gives, exactly,
the cavity equation
\begin{equation}\label{eq:cavity}
\left(G_{jj}\right)^{-1}=\lambda-Y_{jj}-\Sigma_j,\qquad
\Sigma_j=\sum_{k,l\ne j}Y_{jk}\,G^{(j)}_{kl}\,Y_{lj},
\end{equation}
where $G^{(j)}$ is the resolvent of the ``cavity'' matrix with row and column $j$ removed. The
key point is that $G^{(j)}$ contains no element of row $j$ and is therefore statistically
independent of the couplings $Y_{jk}$ that enter $\Sigma_j$. Averaging over these couplings at
fixed $G^{(j)}$, only the terms $k=l$ survive by virtue of Eq. \eqref{eq:Ycov}. This procedure gives
$\frac1N\sum_kS(E_j,E_k)\,G^{(j)}_{kk}$. The quadratic form in the independent couplings $Y_{jk}$ self-averages: at fixed $\eta>0$ its
fluctuations about this mean vanish as $N\to\infty$ (Appendix~\ref{app:cavity}). Note that the limits ($N \to \infty$) and ($\eta \to 0$) do not commute. We take ($N \to \infty$) first, as appropriate for describing the physical properties of the model on scales larger than the level spacing.
Removing a single level changes the sum only by $O\big(1/(N\eta)\big)$, so $G^{(j)}_{kk}$ may be
replaced by $G_{kk}$. Since the denominator of \eqref{eq:cavity} stays at distance $\ge\eta$ from
zero, these errors are amplified at most by $\eta^{-2}$, which is harmless at fixed $\eta$ as
$N\to\infty$. We obtain a closed deterministic system of equations,
\begin{equation}\label{eq:gj}
g_j=\Big[\lambda-\frac1N\sum_kS(E_j,E_k)\,g_k\Big]^{-1}.
\end{equation}
The bath randomness has disappeared, and the Hamiltonian enters only through the profile
$S(E_j,E_k)$. The estimates behind these statements are given in Appendix~\ref{app:cavity}.
Finally, $g_j=g(E_j,\lambda)$ depends on $j$ only through $E_j$, and for $N\to\infty$ the sum over
the levels becomes an integral over the density of states $\nu(E)$. This gives
\begin{equation}\label{eq:dyson}
\begin{aligned}
g(E,\lambda)&=\Big[\lambda-\int\dd E'\,\nu(E')\,S(E,E')\,g(E',\lambda)\Big]^{-1},\\
\rst(\beta)&=-\frac1\pi\,\mathrm{Im}\!\int\dd E\,\nu(E)\,g(E,b+i\eta), \quad \eta \rightarrow 0.
\end{aligned}
\end{equation}
Equation \eqref{eq:dyson} is the vector Dyson equation of a Gaussian matrix with variance profile
$S$~\cite{Shlyakhtenko1996,Ajanki2017,AjankiQVE2019}, which has a unique solution with $\mathrm{Im}\,g<0$ for
$\mathrm{Im}\,\lambda>0$.

\subsection{Universality}\label{sec:universal}
Equation~\eqref{eq:dyson} contains $H_0$ only through its density of states. Two conditions are needed: $H_0$ must be independent of
the bath couplings, and the Lorentzian window must contain many levels, $N\nu(E)\,s\gg1$, so that the
sum over levels in \eqref{eq:gj} may be replaced by an integral. Under these conditions the
eigenvectors of $H_0$, its level correlations (Wigner--Dyson or
Poisson), and its symmetry class (with U(1) symmetry) all drop out, as they did for the width in
Sec.~\ref{subsec:width}. In particular, a diagonal Hamiltonian $H_0=\mathrm{diag}(E_1,\dots,E_N)$
represents every Hamiltonian with the same levels.

\begin{figure}[t]
\centering
\includegraphics{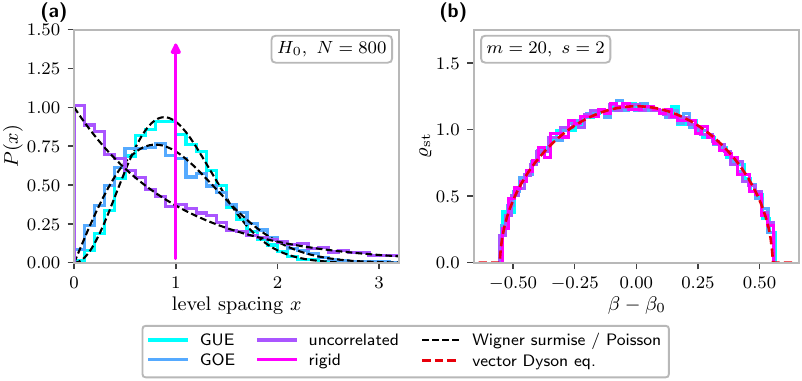}
\caption{Only the density of states matters. (a) Spacings of the unfolded levels in the bulk of four
Hamiltonians with the same semicircular $\nu$ but different level statistics, $N=800$. Dashed:
Wigner surmises and the Poisson law. The rigid levels have all spacings equal to one. (b) $\rst$ from
exact solutions of the Lyapunov equation \eqref{eq:lyap} for the same four Hamiltonians ($N=800$,
$m=20$, $s=2$, $\gamma/\gt=3$), against the vector Dyson equation \eqref{eq:dyson} (red, dashed).}
\label{fig:univ}
\end{figure}
Figure~\ref{fig:univ} tests the statement that only $\nu$ matters, by comparing exact numerical
solutions of Eq.~\eqref{eq:lyap} for Hamiltonians with very different level statistics. For each
realisation we draw $H_0$ and the couplings $\mu,\tilde\mu$, solve the Lyapunov equation directly for $N$ up to $2000$, and diagonalise $\Fst$. We compare the GUE and the GOE, chaotic Hamiltonians with complex and
with real eigenvectors and with quadratic and linear level repulsion, with two diagonal Hamiltonians
of the same semicircular $\nu$: uncorrelated levels, $E_j$ drawn independently from
$\rsc$, which have the Poisson statistics of integrable systems, and rigid levels, $E_j$ placed
at the quantiles of $\rsc$, which are equally spaced in the integrated density. Their level
statistics differ as much as level statistics can [Fig.~\ref{fig:univ}(a)], yet they give the same
$\rst$ [Fig.~\ref{fig:univ}(b)].

\subsection{Closed form of $\rst$}\label{sec:closed}

For the density of states with a bounded support of width $W$, such as the semicircle
\eqref{eq:nusc}, Eq.~\eqref{eq:dyson} can be solved in closed form in two limits.

\subsubsection{\texorpdfstring{$s\gg W$}{s >> W}: a semicircle}\label{sec:larges} 
\begin{figure}[t]
\centering
\includegraphics{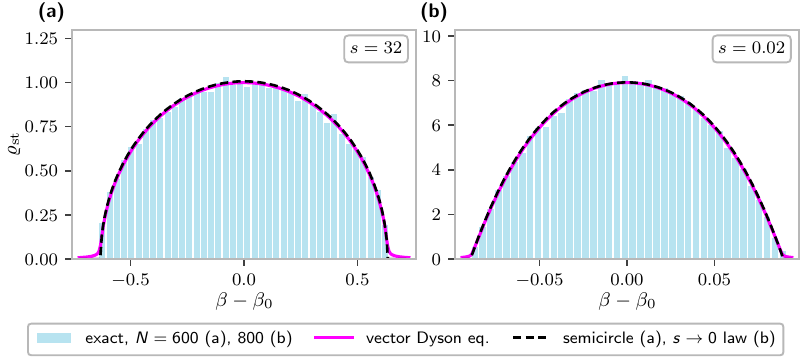}
\caption{The limiting laws in their regimes, at $m=20$. (a) $s=32$ ($\gamma=24$, $\gt=8$, $N=600$,
$5$ realisations) against the semicircle \eqref{eq:larges}. (b) $s=0.02$ ($\gamma=0.015$,
$\gt=0.005$, $N=800$, $4$ realisations) against \eqref{eq:smalls}. Histograms: exact solutions. Solid
line: vector Dyson equation, evaluated at $\eta=\sigma_\beta/60$. Dashed: the limiting law.}
\label{fig:limits}
\end{figure}

When the parameter $s$ exceeds the width of the spectrum, $|E-E'|\le W\ll s$ for every pair of levels, and the profile
\eqref{eq:Sdef} can be expanded in $(E-E')^2/s^2$:
\begin{equation}\label{eq:Slarge}
S(E,E')=\frac{2/m}{1+(E-E')^2/s^2}=\frac2m\Big[1-\frac{(E-E')^2}{s^2}+O(s^{-4})\Big].
\end{equation}
At leading order $S=2/m$ is a constant implying that every level is coupled to every
other one with the same strength. The integral in Eq. \eqref{eq:dyson} then no longer depends on $E$,
\begin{equation}
\int\dd E'\,\nu(E')\,S(E,E')\,g(E',\lambda)=\frac2m\,\avE{g},\qquad
\avE{g}=\int\dd E\,\nu(E)\,g(E,\lambda),
\end{equation}
so that $g(E,\lambda)=\avE{g}$ is independent of $E$ as well. Averaging Eq. \eqref{eq:dyson}
over $E$ with the weight $\nu(E)$ gives a closed scalar equation,
\begin{equation}\label{eq:scalar}
\avE{g}=\frac{1}{\lambda-\frac2m\,\avE{g}}
\qquad\Longleftrightarrow\qquad
\frac2m\,\avE{g}^2-\lambda\,\avE{g}+1=0 .
\end{equation}
Its solution is
\begin{equation}
\avE{g}=\frac m4\Big(\lambda-\sqrt{\lambda^2-8/m}\Big),
\end{equation}
where the branch of the square root is fixed by $\avE{g}\approx 1/\lambda$ at $|\lambda|\to\infty$
(normalisation) and by $\mathrm{Im}\,\avE{g}<0$ for $\mathrm{Im}\,\lambda>0$. On the real axis,
$\lambda=b+i0$, 
for $|b|<2\sqrt{2/m}$, this branch gives $\sqrt{\lambda^2-8/m}=i\sqrt{8/m-b^2}$, hence
$\rst=-\frac1\pi\mathrm{Im}\,\avE{g}=(m/4\pi)\sqrt{8/m-b^2}$, while for $|b|>2\sqrt{2/m}$ the root is
real and $\rst=0$. This results in the semicircle profile
\begin{equation}\label{eq:larges}
\rst(\beta)=\frac{2}{\pi R^2}\sqrt{R^2-b^2},\qquad R=2\sqrt{2/m},
\end{equation}
of variance $R^2/4=2/m$. The same result follows directly at the matrix level: for $s\gg W$ the
factors $r_{jk}$ tend to $1$, so $Y\to X=\xi-\xit$. By \eqref{eq:2pt} the
entries of $X$ have variance $2/M$ and, to leading order in $1/m$, are Gaussian, so $X$ is a Wigner
matrix whose semicircle has variance $N\cdot2/M=2/m$. The result does not depend on $\nu$ as long as its support is bounded. Figure~\ref{fig:limits}(a)
compares it with exact solutions at $s=32$.
\subsubsection{\texorpdfstring{$s\ll W$}{s << W}: a superposition of local semicircles}\label{sec:smalls}
\begin{figure}[t]
\centering
\includegraphics{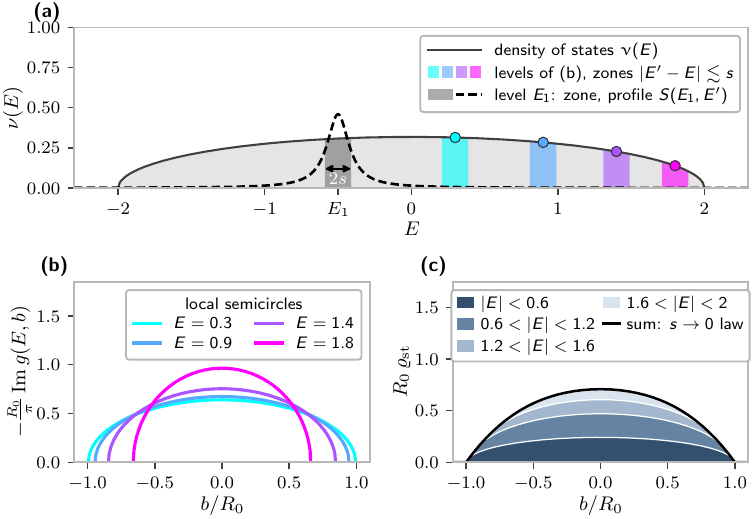}
\caption{Local semicircles at $s\ll W$ for the GUE. (a) Density of states $\nu(E)$. Coloured: four levels
(circles) and their neighbours within $\sim s$, which give the bulk of the coupling $u(E)$. The area of
each strip is $\propto\nu(E)\propto R(E)^2$. Grey and dashed: the coupling zone and the coupling profile
$S(E_1,E')$ of a level $E_1$. The width $s$ is exaggerated. (b) Local semicircles $-\frac1\pi\mathrm{Im}\,g(E,b)$ of
the four levels marked in (a). Their radii $R(E)=R_0(1-E^2/4)^{1/4}$ shrink towards the band edges.
(c) The $s\to0$ law \eqref{eq:smallgen} as the sum of the local semicircles weighted with $\nu(E)$,
shaded by the band of $|E|$ they come from (grey-blue, unrelated to the colours of the levels in
(a) and (b)). Only the levels near the band centre reach the edge
$b=R_0$. The axes in (b) and (c) are in units of $R_0=2\sqrt{2s/m}$.}
\label{fig:locsc}
\end{figure}

When $s\ll W$, the Lorentzian in \eqref{eq:Sdef} is narrow compared with the scale on which
$\nu(E')$ and $g(E',\lambda)$ vary. Since $\frac1\pi\frac{s}{x^2+s^2}\to\delta(x)$ as $s\to0$, 
the profile becomes a diagonal kernel,
\begin{equation}\label{eq:Ssmall}
S(E,E')\;\longrightarrow\;\frac{2\pi s}{m}\,\delta(E-E').
\end{equation}
This requires the continuum limit to be taken first: the Lorentzian must still cover many levels,
$N\nu(E)\,s\gg1$. Inserted in Eq. \eqref{eq:dyson}, the integral becomes local in energy,
\begin{equation}
\int\dd E'\,\nu(E')\,S(E,E')\,g(E',\lambda)=u(E)\,g(E,\lambda),\qquad
u(E)=\frac{2\pi s}{m}\,\nu(E),
\end{equation}
where $u(E)$ is the sum rule \eqref{eq:sumrule} in this limit. The equations for different $E$
decouple. At each $E$ one obtains the scalar equation \eqref{eq:scalar} with $2/m\to u(E)$,
\begin{equation}\label{eq:localg}
g(E,\lambda)=\frac{1}{\lambda-u(E)\,g(E,\lambda)},
\end{equation}
and its solution is taken over from Sec.~\ref{sec:larges}: a semicircle
\begin{equation}
-\frac1\pi\,\mathrm{Im}\,g(E,b+i0)=\frac{2\sqrt{R(E)^2-b^2}}{\pi R(E)^2}\,\theta\big(R(E)-|b|\big),
\qquad R(E)^2=4u(E)=\frac{8\pi s}{m}\,\nu(E),
\end{equation}
whose radius depends on the position in the band. Figure~\ref{fig:locsc} illustrates this picture
for the GUE. A level at energy $E$ is coupled by $Y$ only to its neighbours within
$|E-E'|\lesssim s$, whose number is proportional to the local density $N\nu(E)$
[Fig.~\ref{fig:locsc}(a)]. It therefore sees a local Wigner matrix of variance $u(E)\propto\nu(E)$,
whose spectrum is a local semicircle [Fig.~\ref{fig:locsc}(b)]. By the second line of
Eq. \eqref{eq:dyson}, $\rst$ is the average of these local densities with the weight $\nu(E)$
[Fig.~\ref{fig:locsc}(c)], and in this average the prefactors combine into a constant,
$\nu(E)\cdot2/[\pi R(E)^2]=m/(4\pi^2s)$. Altogether,
\begin{equation}\label{eq:smallgen}
\rst(\beta)=\frac{m}{4\pi^2s}\int\dd E\,\big[R(E)^2-b^2\big]_+^{1/2},\qquad
\sigma_\beta^2=\frac{2\pi s}{m}\int\dd E\,\nu(E)^2 ,
\end{equation}
where $[x]_+=\max(x,0)$. The variance agrees with the small-$s$ limit of \eqref{eq:psidef}. The
support ends at the largest local radius, $b_e=\max_ER(E)$, i.e.\ at the maximum of the
density of states. Only the levels near this maximum reach the edge [Fig.~\ref{fig:locsc}(c)].

\begin{figure}[t]
\centering
\includegraphics{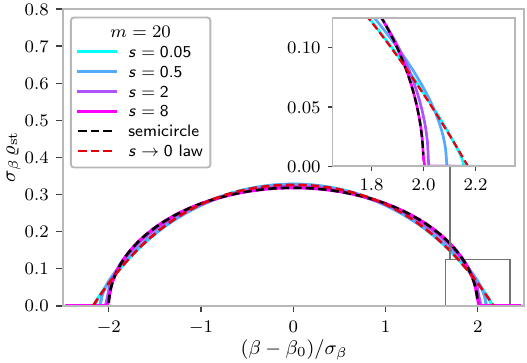}
\caption{Crossover of the shape. The vector Dyson solution at $m = 20$ for $s =$ 0.05,
0.5, 2 and 8, rescaled to unit variance (colours), with the semicircle (black, dashed) and
the $s\rightarrow 0$ law \eqref{eq:localsc} (red, dashed). The inset magnifies the right edge, marked
by the box.}
\label{fig:crossover}
\end{figure}

For the GUE the local radius is $R(E)^2=R_0^2\,(1-E^2/4)^{1/2}$, with the largest radius
$R_0=R(0)=2\sqrt{2s/m}$, and the prefactor is $m/(4\pi^2s)=2/(\pi^2R_0^2)$. The radius is largest at the band centre and vanishes at the band edges.
The step function restricts the integration to $R_0^2(1-E^2/4)^{1/2}\ge b^2$, i.e.\ $E^2\le E_*^2$ with
$E_*^2=4(1-b^4/R_0^4)$, and
\begin{equation}\label{eq:localsc}
\rst(\beta)=\frac2{\pi^2R_0^2}\int_{-E_*}^{E_*}\dd E\,\Big[R_0^2\Big(1-\frac{E^2}4\Big)^{1/2}-b^2\Big]^{1/2},
\end{equation}
which is non-zero for $|b|\le R_0$. The integral can be computed explicitly:
\begin{equation}\label{eq:smalls}
\rst(\beta)=\frac{16\sqrt{R_0^2+b^2}}{3\pi^2R_0^4}\Big[R_0^2\,\Kel(p)-b^2\,\Eel(p)\Big],\qquad
p=\frac{b^2-R_0^2}{b^2+R_0^2},\qquad |b|\le R_0 ,
\end{equation}
where we recall that $R_0=R(0)=2\sqrt{2s/m}$.
The density $\rst$ is normalised, with
\begin{equation}\label{eq:smallsmom}
\sigma_\beta^2=\frac{16s}{3\pi m},\qquad
\frac{\av{b^4}}{\av{b^2}^2}=\frac{27\pi^2}{128} .
\end{equation}
The variance agrees with the small-$s$ asymptote in \eqref{eq:asym}.  The fourth-moment ratio should be compared with $2$ for a
semicircle. Two further signatures distinguish \eqref{eq:smalls} from a semicircle. The edge of the limiting law
is \emph{linear}, $\rst\propto R_0-|b|$, and it lies at
$R_0=\sqrt{3\pi/2}\,\sigma_\beta>2\sigma_\beta$, i.e.\ further out in units of the width than
the semicircle edge. That linear edge, however, belongs to the $s\to0$ law and not to any finite
$s$: as shown in Sec.~\ref{sec:edge}, at every $s>0$ the true edge is a square root, confined to a
window $\delta=b_e-b\sim s\,\sigma_\beta$ which closes as $s\to0$. Outside it the density is again linear
and approaches the straight line above for $R_0-b\gg sR_0\ln(1/s)$. Figure~\ref{fig:limits}(b) compares \eqref{eq:smalls} with exact solutions
at $s=0.02$, and Fig.~\ref{fig:crossover} shows how the shape crosses over between the two limiting
laws as $s$ grows.

\section{The edge of the distribution}\label{sec:edge}
The edge $b_e$ of $\rst$ gives, at $N,m\to\infty$, the largest modular energy,
$\beta_{\max}\simeq\beta_0+b_e$, i.e.\ the least occupied natural orbital. All results of this section
refer to the edge of the leading-order density, with $m\to\infty$ taken at fixed $s$. For finite
matrices the largest $\beta_j$ fluctuates about this value, and the corrections of order $1/m$ to $\Hst$
must be small on the scale of the window discussed below. Let $\delta=b_e-b$ be the distance from the edge. The two limiting
laws of Sec.~\ref{sec:closed} vanish differently there: the semicircle \eqref{eq:larges} as
$\sqrt\delta$, the $s\to0$ law \eqref{eq:smalls} linearly. We show that at every $s>0$ the edge is a
square root, $\rst\simeq A\sqrt\delta$ with $A>0$. At small $s$ it holds only within a window
$\delta\lesssim s\,\sigma_\beta$, beyond which the density approaches the linear law, so that the
limits $\delta\to0$ and $s\to0$ do not commute. The limit $N\to\infty$ is taken first, and $s\to0$
means $\Delta\ll s\ll W$, with $\Delta=1/(N\nu)$ the level spacing of $H_0$. The results are collected
in Table~\ref{tab:edge} and compared with the numerical solution of the vector Dyson equation in
Fig.~\ref{fig:edge}. The derivations are given in Appendix~\ref{app:edge}.

\begin{figure}[t]
\centering
\includegraphics{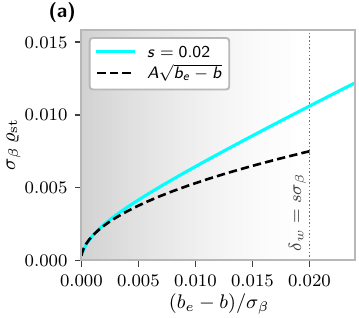}\hspace{1.2em}
\includegraphics{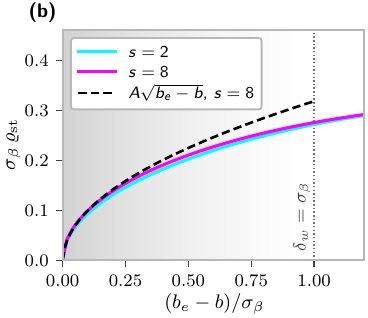}
\caption{The density near the upper edge, in units of $\sigma_\beta$. Solid: numerical solution of
\eqref{eq:Top}. Dashed: $A\sqrt\delta$, $\delta=b_e-b$, with $b_e$ and $A$ from \eqref{eq:small} in (a) and
from \eqref{eq:lsprofiles}--\eqref{eq:lsA} for $s=8$ in (b). Grey: the window of the square root, up to
$\delta_w$ (dotted), with $\delta_w=s\,\sigma_\beta$ in (a) and $\sigma_\beta$ in (b). (a) $s=0.02$: beyond the
window the density bends towards the linear law \eqref{eq:smalls}. (b) $s=2$ and $8$: the deviation
from the dashed line is the curvature of the semicircle.}
\label{fig:edge}
\end{figure}

\begin{table}[t]
\centering
\begin{tabular}{lcc}
\toprule
 & $b_e/\sigma_\beta$ & $A\,\sigma_\beta^{3/2}$\\
\midrule
$s\gg W$ & $2+O(s^{-4})$ & $1/\pi+O(s^{-4})$\\
$s\ll W$ & $\sqrt{3\pi/2}\,\big[1-O(s\ln\frac1s)\big]$ & $0.38\sqrt s$\\
\bottomrule
\end{tabular}
\caption{Position $b_e$ and amplitude $A$ of the square-root edge $\rst\simeq A\sqrt{b_e-b}$ of the
GUE in units of the width $\sigma_\beta$, Eqs.~\eqref{eq:lsprofiles}, \eqref{eq:lsA},
\eqref{eq:small} and \eqref{eq:Ahat}. The coefficient $0.38$ follows from the numerical solution of
Eq. \eqref{eq:universal}.}
\label{tab:edge}
\end{table}

\subsection{The square-root law}\label{sec:sqrt}
We write the vector Dyson equation \eqref{eq:dyson} as
\begin{equation}\label{eq:Top}
\begin{aligned}
g(E,\lambda)&=\frac{1}{\lambda-\TT[g](E)},\qquad
&\TT[f](E)&=\int\dd E'\,\nu(E')\,S(E,E')\,f(E'),\\[2pt]
\rst&=-\frac1\pi\,\mathrm{Im}\,\avE{g(\cdot,b+i0)},
&\avE{f}&=\int\dd E\,\nu(E)\,f(E),
\end{aligned}
\end{equation}
where $\avE{\,\cdot\,}$ is an average over the band, not over disorder. Since $S(E,E')=S(E',E)$, the
operator $\TT$ is symmetric. In this subsection $\nu$ is arbitrary.

Let $g=g_r-ig_i$, so that $\rst=\avE{g_i}/\pi$. Outside the support $g_i=0$. The density appears at
the point $b_e$ where the equation for $g_i$, linearised about the real solution, acquires a
non-zero solution $w(E)$:
\begin{equation}\label{eq:thresh}
w=g_e^2\,\TT[w],\qquad g_e(E)=g_r(E,b_e),\qquad \frac1{g_e}=b_e-\TT[g_e].
\end{equation}
This is an eigenvalue problem. Far from the band $g_r\simeq1/b$ is small, and the edge is the point
at which the largest eigenvalue of the positive operator $g_r^2\TT$ reaches one. By the
Perron--Frobenius theorem its eigenfunction is positive, $w>0$. It is the shape of $g_i$ across the
band near the edge. The next order of the expansion in $g_i$ fixes its amplitude
(Appendix~\ref{app:sqrt}) and gives
\begin{equation}\label{eq:root}
\rst(\beta)=A\,\sqrt\delta\,\big[1+O(\delta/\delta_w)\big],\qquad
A=\frac{\avE{w}^{3/2}}{\pi\,\avE{w^3/g_e^3}^{1/2}} .
\end{equation}
Both averages are positive, so the edge is a square root with a non-zero amplitude at every $s>0$.
The higher orders in $\delta$ set the width $\delta_w$ of the window in which it holds. For a flat
profile, $S=2/m$, one has $w=1$, $g_e=\sqrt{m/2}$, $b_e=2\sqrt{2/m}$ and $A=(m/2)^{3/4}/\pi$, the
edge of the semicircle \eqref{eq:larges}.

\subsection{\texorpdfstring{$s\gg W$}{s >> W}: a rescaled semicircle}\label{sec:lsedge}
At $s\gg W$ the profile \eqref{eq:Slarge} is flat up to terms of order $s^{-2}$. A level at energy
$E$ then feels the band only through $\avE{g}$, with the coupling $u(E)$ of the sum rule
\eqref{eq:sumrule}. Averaging over the band gives the scalar equation \eqref{eq:scalar} with
$2/m\to\avE{u}=\sigma_\beta^2$, up to relative corrections of order $s^{-4}$
(Appendix~\ref{app:lsedge}). The solution is the semicircle of variance $\sigma_\beta^2$, and to first
order in $s^{-2}$
\begin{equation}\label{eq:lsprofiles}
b_e\simeq2\sigma_\beta\simeq2\sqrt{\frac2m}\left(1-\frac{\sigma_\nu^2}{s^2}\right),
\end{equation}
\begin{equation}\label{eq:lsA}
A\simeq\frac{1}{\pi\,\sigma_\beta^{3/2}}\simeq\frac1\pi\Big(\frac m2\Big)^{3/4}\left(1+\frac{3\sigma_\nu^2}{2s^2}\right),
\end{equation}
where $\sigma_\nu^2$ is the variance of $\nu$, equal to $1$ for the GUE. In units of the width the
edge and the amplitude keep their semicircle values, $b_e/\sigma_\beta=2$ and
$A\sigma_\beta^{3/2}=1/\pi$: the first correction only rescales the semicircle. The square root holds
for $\delta\ll b_e$, i.e.\ $\delta_w\sim\sigma_\beta$.

\subsection{\texorpdfstring{$s\ll W$}{s << W}: a boundary layer at the band centre}\label{sec:layer}

In this subsection we consider $H_0\in$ GUE only, such that $\nu=\rsc$. At $s\to0$ every level $E$ of $H_0$ contributes its own
local semicircle of radius $R(E)$, Eq.~\eqref{eq:smallgen}. The radius is largest at the band centre, $R_0$,
where the levels are densest, and it decreases away from it,
$R(E)\simeq R_0(1-E^2/16)$. In units of the width $R_0\simeq\sqrt{3\pi/2}\,\sigma_\beta$. Just below $R_0$
only the levels whose semicircles still reach $b$ contribute. These are the levels near the centre,
$E^2<16(R_0-b)/R_0$: as $b\to R_0$ the whole edge of $\rst$ is made by the band centre alone. Each of
these levels sits close to its own edge and contributes a local density
$\propto[R(E)-b]^{1/2}\lesssim(R_0-b)^{1/2}$, and their number is proportional to the width
$\propto(R_0-b)^{1/2}$ of the window in $E$. The product of the two square roots is linear in $R_0-b$,
which is the linear edge of Eq.~\eqref{eq:smalls}. At finite $s$ the levels are coupled through the Lorentzian profile $S$. Its slowly decaying tail,
$\propto s^2/(E-E')^2$, acts on a smooth solution as a nonlocal operator of first order in $s$. As
$b\to R_0$ the window of contributing levels shrinks, and once it is of order $\sqrt s$ this coupling
is no longer negligible. The levels near the centre then form one random matrix with one common edge,
which as shown in Sec.~\ref{sec:sqrt} is a square root. The derivation is given in Appendix~\ref{app:layer}. Its outline is as follows.

Near the centre the local coupling is $u(E)\simeq u_0(1-E^2/8)$, $u_0=u(0)\simeq(2s/m)(1-s/2)$, and the
edge of the central level is at $b=2\sqrt{u_0}$. The coupling between the levels acts on a function
that varies slowly on the scale $s$ as $s\,|\partial_E|$, where $|\partial_E|$ is the operator with
the Fourier symbol $|k|$, Eq.~\eqref{eq:absd}. It is of first order in $s$, and away from the centre
it may be dropped. This gives the picture of independent local semicircles and the linear edge. Near
the centre, however, the coupling is comparable to the decrease $E^2/8$ of the local coupling in a
region $|E|\lesssim\sqrt s$, however small $s$ is. This region is a boundary layer, of the kind solved
by matched asymptotic expansions~\cite{BenderOrszag1999}. In the stretched variables $E=\sqrt s\,y$
and $g=(1+\sqrt s\,\psi)/\sqrt{u_0}$, and with $b=2\sqrt{u_0}\,(1-s\,q)$, the vector Dyson equation
near the edge reduces to
\begin{equation}\label{eq:universal}
\psi^2-|\partial_y|_{\rm reg}\,\psi=\frac{y^2}{8}-2Q,\qquad
Q=q-\frac{\ln(1/s)}{4\sqrt2\,\pi}-\frac{C_{\rm out}}2 ,
\end{equation}
where $|\partial_y|_{\rm reg}$ is $|\partial_y|$ with the logarithm from the distant levels subtracted,
Eq.~\eqref{eq:dreg}. The levels outside the layer shift the distance from the edge $Q$ against $q$ by
the logarithm and by the constant
$C_{\rm out}=-\frac12+[1+3\ln2-\ln(1+\sqrt2)]/(\pi\sqrt2)$, Eq.~\eqref{eq:Coutval}.
As an equation for $\psi(y)$ at given $Q$, Eq.~\eqref{eq:universal} contains neither $s$ nor a cutoff. The dependence on $s$ enters only through the relation between $Q$ and $b$. It is a stationary
Benjamin--Ono equation with a quadratic potential, and we solve it numerically. Outside the spectrum
its solution is real. The edge $Q_*\approx0.0276$ is the largest $Q$ at which a real solution exists:
there two real solutions merge, as in the scalar equation, and the linearised equation has the
positive solution $w(y)$, the edge mode of Sec.~\ref{sec:sqrt}. Inserting it in \eqref{eq:root} gives
the amplitude through the number $a_0\approx0.254$, Eq.~\eqref{eq:layerA}. With
$2\sqrt{u_0}\simeq R_0\,(1-s/4)$ the edge and its amplitude are
\begin{equation}\label{eq:small}
\begin{gathered}
b_e=2\sqrt{\frac{2s}{m}}\left[1-\frac{s}{\sqrt2}\Big(c_0+\frac{1}{4\pi}\ln\frac1s\Big)\right],
\qquad A=a_0\,\frac{m^{3/4}}{s^{1/4}},\\
c_0=\frac1{2\sqrt2}+\sqrt2\Big(Q_*+\frac{C_{\rm out}}2\Big)=\frac{1+3\ln2-\ln(1+\sqrt2)}{2\pi}+\sqrt2\,Q_* \simeq 0.39.
\end{gathered}
\end{equation}

The correction of order $s$ to the edge in \eqref{eq:small} has three sources. The reduction of the
coupling of the central level at finite $s$, Eq.~\eqref{eq:u}, gives the term $1/(2\sqrt2)$ in $c_0$. The
distant levels give the logarithm and the constant $C_{\rm out}$. These two contributions are known in
closed form and combine into $[1+3\ln2-\ln(1+\sqrt2)]/2\pi$. The layer adds
$\sqrt2\,Q_*$. The constant $c_0$ is not small against the logarithm, which enters with the
coefficient $1/4\pi$: at $s=0.02$ the two are $0.39$ and $0.31$. The layer thus shifts the edge only
slightly, but it alone fixes the amplitude $a_0$ and the shape of the edge. Both $c_0$ and $a_0$ agree
with the $s\to0$ extrapolation of the numerical solution of \eqref{eq:Top}, obtained by a Newton
iteration on a grid in $E$. Figure~\ref{fig:edge} compares Eqs.~\eqref{eq:lsprofiles}--\eqref{eq:lsA}
and \eqref{eq:small} with this solution, and no parameter is fitted to the curves.

In units of the width, the amplitude vanishes,
\begin{equation}\label{eq:Ahat}
A\,\sigma_\beta^{3/2}=a_0\Big(\frac{16}{3\pi}\Big)^{3/4}\sqrt s \xrightarrow{\ s\to0\ }\ 0 ,
\end{equation}
with $\sigma_\beta=(16s/3\pi m)^{1/2}$, Eq.~\eqref{eq:smallsmom}. The amplitude $A$ itself grows as $s\to0$
only because $\rst$ is normalised while its width shrinks. The same inner and outer structure appears
in the density. The square root is the inner region, $\delta\lesssim s\,\sigma_\beta$. Beyond it,
$\delta\gg s\,\sigma_\beta$, the density becomes linear again, but with the edge at $b_e$. The edge of the
linear law \eqref{eq:outerrho} of independent levels lies above $b_e$ by $O(s\ln\frac1s)\,R_0$, so the
density approaches this unshifted law only for $R_0-b\gg sR_0\ln\frac1s$.

At finite $N$ the level spacing returns. The square-root window $\delta\sim s\,\sigma_\beta$ contains
about $NA\,\delta^{3/2}\sim Ns^2$ eigenvalues of $Y$, so the square-root edge is resolved only for
$Ns^2\gg1$, a stronger condition than $Ns\gg1$. For $s\lesssim\Delta$ the Lorentzian no longer couples
a level to its neighbours, $Y$ approaches the diagonal matrix of the $X'_{jj}$, and $\rst$ narrows to
a Gaussian of variance $2/M$. The corrections then come from close pairs of levels,
$|E_j-E_k|\lesssim s$, whose number depends on the level statistics of $H_0$, and the universality of
Sec.~\ref{sec:universal} no longer holds.

\section{Confinement of the modular energies and L\'evy Hamiltonians}\label{sec:confine}
\subsection{A bound for every Hamiltonian}\label{subsec:bound}
In this section we answer the question: how far can a single mode be driven from $\beta_0$? The row sum of the variance profile satisfies
$u(E)\le 2/m$ for any normalised density of states. In the
leading Gaussian approximation, the limiting spectral density
of $Y$ is therefore supported within
\begin{equation}\label{eq:confine}
\operatorname{supp}\rho_Y\subseteq
[-2\sqrt{u_{\max}},2\sqrt{u_{\max}}]
\subseteq[-2\sqrt{2/m},2\sqrt{2/m}],
\qquad u_{\max}=\sup_Eu(E).
\end{equation}
Here $\rho_Y(b)$ gives the leading approximation to
$\rst(\beta_0+b)$. The outer bound is saturated in this
approximation when the profile becomes constant, as for
$H_0=0$ or $s\gg W$ with bounded spectral support.

There is also an exact contraction statement for the linearised
matrix, valid for every realisation and every finite $N$:
\begin{equation}\label{eq:exactYbound}
\|Y\|\le s\int_0^\infty\dd t\,e^{-st}\|X(t)\|=\|X\|.
\end{equation}
The deterministic number $2\sqrt{2/m}$ replaces $\|X\|$ only
at leading order in the large-$N$, large-$m$ limit. It is not
an exact bound on individual realisations or on the exact
modular Hamiltonian at finite $m$.
A separate exact bound is available for the modular energies.
In an eigenbasis of $F_{\rm st}$, the diagonal of
$[H_0,F_{\rm st}]$ vanishes. Equation~\eqref{eq:lyap} then gives
\begin{equation}\label{eq:exactRateRatio}
f_j=\frac{\Gamma_{jj}-\Gt_{jj}}
{\Gamma_{jj}+\Gt_{jj}},\qquad
\beta_j=\ln\frac{\Gamma_{jj}}{\Gt_{jj}}.
\end{equation}
For positive definite dissipators, the Rayleigh bounds imply
\begin{equation}\label{eq:exactBetaBound}
\ln\frac{\lambda_{\min}(\Gamma)}{\lambda_{\max}(\Gt)}
\le\beta_j\le
\ln\frac{\lambda_{\max}(\Gamma)}{\lambda_{\min}(\Gt)} ,
\end{equation}
where $\lambda_{\min}(A)$ and $\lambda_{\max}(A)$ are the smallest and largest eigenvalues of the matrix $A$.
This bound holds for every $H_0$ and every realisation, but
its endpoints depend on the realised bath matrices. For the
Wishart baths at fixed $m>1$, their large-$N$ spectral edges
give the weaker asymptotic estimate
\begin{equation}\label{eq:MPBetaBound}
\max_j|\beta_j-\beta_0|
\le 4\artanh(m^{-1/2})+o_N(1).
\end{equation}
Equation~\eqref{eq:MPBetaBound} concerns the exact modular
energies, whereas Eq.~\eqref{eq:confine} gives the sharper
support estimate within the leading large-$m$ theory.

\subsection{L\'evy Hamiltonians}\label{subsec:levy}
The bound \eqref{eq:confine} is tested most sharply by Hamiltonians with an unbounded, heavy-tailed
spectrum and eigenstates localised in the spectral tails. We take L\'evy matrices~\cite{Cizeau1994,BenArous2008}: real
symmetric, with independent upper-triangular entries $x=(H_0)_{ij}$ distributed as $P(x)=h^\alpha/|x|^{1+\alpha}$,
$h=N^{-1/\alpha}$, $0<\alpha<2$, as in Ref.~\cite{Cizeau1994}. The distribution vanishes for
$|x|<(2/\alpha)^{1/\alpha}h$, which normalises it. Their density of states has no finite band and a power-law tail,
$\nu(E)\simeq c_\alpha|E|^{-1-\alpha}$, since the largest eigenvalues are the largest entries. Below
$c_\alpha$ is the amplitude of such a tail for an arbitrary $H_0$, and the formulas use only this tail.
For the L\'evy matrices above $c_\alpha^{\rm(L)}=Nh^\alpha=1$. The density of states is known
exactly~\cite{BenArous2008,Safonova2025},
\begin{equation}\label{eq:levydos}
\nu(E)=\mathrm{Re}\,\frac1{2\pi}\int_0^\infty\dd y\;e^{\frac i2Ey-\zeta y^{\alpha/2}},\qquad
\zeta=\frac{\Gamma(1-\frac\alpha2)}{2^\alpha\,\Gamma(1+\frac\alpha2)}\int_0^\infty\dd y\;y^{\frac\alpha2-1}\,
e^{\frac i2Ey-\zeta y^{\alpha/2}},
\end{equation}
where $\zeta(E)$ is found self-consistently. This is the result of Ref.~\cite{Safonova2025}, where
the entries are drawn from the stable law $e^{-|k|^\alpha}$, rewritten for the normalisation above.
Only the prefactor of the second equation changes. The cavity approach of Ref.~\cite{Cizeau1994}
leads to the same equations if the asymmetry of the stable law of the cavity self-energy is weighted
with $|G|^{\alpha/2}$, where $G$ is a diagonal element of the resolvent of $H_0$. A term of the
self-energy contributes to the tail on its side in proportion to
$|G|^{\alpha/2}$. Ref.~\cite{Cizeau1994} uses the unweighted asymmetry $\langle\mathrm{sgn}\,G\rangle$.
This gives the same $\nu(0)$ and the same tail, but changes $\nu$ in the bulk. Equation~\eqref{eq:levydos} agrees with direct diagonalisation.

\begin{figure}[t]
\centering
\includegraphics{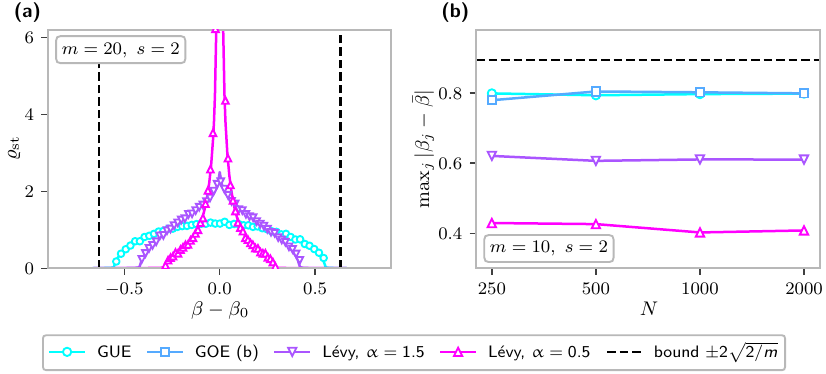}
\caption{Confinement for heavy-tailed Hamiltonians. (a) $\rst$ for GUE and L\'evy $H_0$ ($m=20$,
$s=2$): exact solutions of \eqref{eq:lyap} (symbols, $10$ realisations with $N=800$ levels) and the vector Dyson
equation \eqref{eq:dyson} with the semicircle (GUE) and with the exact L\'evy density of states
\eqref{eq:levydos} (lines). Dashed: $\pm2\sqrt{2/m}$. (b) Largest $|\beta_j-\bar\beta|$ against $N$
($m=10$, $s=2$) and the bound \eqref{eq:confine} (dashed). Over the same samples the largest
eigenvalue of the L\'evy $H_0$ ranges from $17$ to $9\cdot10^8$.}
\label{fig:compact}
\end{figure}

Figure~\ref{fig:compact}(b) shows the largest $|\beta_j-\bar\beta|$ from exact solutions of the
Lyapunov equation. The largest eigenvalues of the L\'evy Hamiltonians grow as $N^{1/\alpha}$, but the
largest $|\beta_j-\bar\beta|$ does not follow them and saturates below the bound \eqref{eq:confine},
as for the GUE and GOE. It is even lower than for the GUE, and $\rst$ is narrower in
Fig.~\ref{fig:compact}(a): most levels of a broad spectrum have few neighbours within $s$, and $u$
is small. Figure~\ref{fig:compact}(a) shows that the vector Dyson equation with the
exact density of states reproduces the exact $\rst$, including a sharp peak at its centre. The heavy
tails leave two signatures: a slow approach to the semicircle and a non-analytic centre.

\subsection{Slow approach to the semicircle}\label{subsec:slow}
As $s\to\infty$ the variance profile tends to a constant, and $\rst$ tends to the semicircle of radius
$2\sqrt{2/m}$. How fast it does so is measured by $1-\Psi$, since $\sigma_\beta^2=2\Psi/m$. For a
bounded spectrum $1-\Psi\simeq2\sigma_\nu^2/s^2$, where $\sigma_\nu^2$ is the variance of $\nu$ ($\sigma_\nu^2=1$ for the GUE, Eq.~\eqref{eq:asym}). For a heavy tail the fraction of level
pairs farther apart than $s$ decays only as a power of $s$, and the approach is slower. Since
$1-\Psi=\big\langle x^2/(x^2+s^2)\big\rangle$ over the separation $x=E-E'$ of two levels drawn from
$\nu$, and the second moment of $x$ is infinite for $\alpha<2$, the average is dominated by
$|x|\sim s$, where the density of $x$ is $2c_\alpha|x|^{-1-\alpha}$. Rescaling $x=sy$,
\begin{equation}\label{eq:onemPsi}
1-\Psi(s)\simeq4c_\alpha s^{-\alpha}\int_0^\infty\dd y\,\frac{y^{1-\alpha}}{1+y^2}
=\frac{2\pi c_\alpha}{\sin(\pi\alpha/2)}\,s^{-\alpha},\qquad 0<\alpha<2 .
\end{equation}
For a Cauchy density of states, $\nu=c_1/(E^2+\pi^2c_1^2)$ with any $c_1$, the exact result
$1-\Psi=2\pi c_1/(2\pi c_1+s)$ reproduces this asymptote.

\subsection{Non-analytic centre}\label{subsec:centre}
A level deep in the tail, $|E|\gg s$, receives its noise mostly from its neighbours within the
Lorentzian window, so its contribution to $\rst$ is a local semicircle, as in
Sec.~\ref{sec:smalls}. Its radius is $R(E)\propto\nu(E)^{1/2}\propto|E|^{-(1+\alpha)/2}$. The coupling to the
bulk through the tail of the Lorentzian broadens this semicircle by an amount $\propto E^{-2}$, which
is small compared with $R(E)$ for $\alpha<3$. Tail levels are rare, but their
semicircles are narrow and therefore tall, with height $2/[\pi R(E)]\propto\nu^{-1/2}$. A level
contributes to $\rst(b)$ only inside its own semicircle, $|b|<R(E)$. Since $R(E)$ decreases along the
tail, the levels that contribute at a given $b$ are those with $|E|$ below some $E_b$, and $E_b\to\infty$
as $b\to0$. The closer $b$ is to the centre, the more of the narrow, tall tail semicircles add up
there. Eq.~\eqref{eq:smallgen} applied to the tail gives
\begin{equation}\label{eq:tail2}
\rst(b)\simeq\rst^{\rm reg}+k_\alpha\Big[\frac{2^{2-\alpha}c_\alpha}{\pi^{1+2\alpha}}\Big(\frac{m}{s}\Big)^{\alpha}\Big]^{\frac1{1+\alpha}}
|b|^{-\frac{1-\alpha}{1+\alpha}},\qquad
k_\alpha=\frac{1}{1+\alpha}\,\mathrm B\Big(\frac{1-\alpha}{2(1+\alpha)},\frac32\Big),
\end{equation}
where $\rst^{\rm reg}$ is the regular part. To obtain Eq. \eqref{eq:tail2}, we insert the tail
$\nu\simeq c_\alpha|E|^{-1-\alpha}$ into Eq. \eqref{eq:smallgen}, so that we find
$R(E)^2=(8\pi s\,c_\alpha/m)|E|^{-1-\alpha}$. A tail level contributes as long as $R(E)>|b|$, i.e.\ up to
$E_b=(8\pi s\,c_\alpha/mb^2)^{1/(1+\alpha)}$, and with $|E|=E_b\,t$ both tails give
\begin{equation*}
\frac{m}{4\pi^2s}\,2\int\dd E\,\big[R(E)^2-b^2\big]_+^{1/2}
=\frac{m}{2\pi^2s}\,E_b\,|b|\int_0^1\dd t\,\big(t^{-1-\alpha}-1\big)^{1/2},
\end{equation*}
with $E_b|b|=(8\pi s\,c_\alpha/m)^{1/(1+\alpha)}|b|^{-(1-\alpha)/(1+\alpha)}$. Changing the
variable to $t^{1+\alpha}$ turns the integral into $k_\alpha$. For $\alpha<1$ it converges at $t\to0$, i.e.\ it is set
by the far tail. For $\alpha>1$ it diverges there, at $|E|\ll E_b$, i.e.\ for levels with $R(E)\gg|b|$.
Their leading contribution, $[R(E)^2-b^2]^{1/2}\simeq R(E)$ in the integrand over $E$, does
not depend on $b$ and belongs to $\rst^{\rm reg}$. Subtracting it leaves the convergent integral
\begin{equation*}
\int_0^1\dd t\,\Big[\big(t^{-1-\alpha}-1\big)^{1/2}-t^{-\frac{1+\alpha}2}\Big]+\frac{2}{1-\alpha}=k_\alpha ,
\end{equation*}
which is the same Beta function, now with a negative first argument (for $\alpha<1$ the subtraction
and the term $2/(1-\alpha)$ cancel).
For $\alpha<1$, $k_\alpha>0$, and $\rst$ has a divergent
peak at the centre. For $1<\alpha<2$, $k_\alpha<0$, but the exponent in Eq. \eqref{eq:tail2} is
then positive: $\rst(0)$ is finite, and the tail term lowers $\rst$ away from $b=0$ as
$-|b|^{(\alpha-1)/(\alpha+1)}$. The centre is therefore a sharp peak, a cusp with infinite slope, as for
$\alpha=1.5$ in Fig.~\ref{fig:compact}(a). At $\alpha=1$ the divergence is logarithmic,
$\rst\simeq\big[2c_\alpha m/(\pi^3s)\big]^{1/2}\ln(1/|b|)$. The coefficients in Eq. \eqref{eq:tail2} agree
with the solution of the vector Dyson equation with the exact density of states. Heavy tails thus
act at the centre of the distribution, not at its edge.

\subsection{The edge for heavy-tailed Hamiltonians}\label{subsec:levyedge}
At small $s$, the leading location of the edge is determined
by the maximum of the Hamiltonian density of states:
\begin{equation}\label{eq:levySmallEdge}
b_e\sim2\sqrt{\frac{2\pi s}{m}\sup_E\nu(E)}.
\end{equation}
For each fixed $0<\alpha<2$, Eq.~\eqref{eq:levydos} gives a
finite, smooth density at $E=0$. In particular, setting
$K_{\alpha/2}=\Gamma(1-\alpha/2)/[2^{\alpha}\Gamma(1+\alpha/2)]$, one obtains
\begin{equation}\label{eq:levyZeroDOS}
\zeta(0)=\sqrt{2K_{\alpha/2}/\alpha},\qquad
\nu(0)=\frac{\Gamma(2/\alpha)}{\pi \alpha}
\left(\frac{\alpha/2}{K_{\alpha/2}}\right)^{1/\alpha}.
\end{equation}
Thus a narrow central peak can delay the onset of the local
small-$s$ regime, but does not change the asymptotic
$b_e\propto\sqrt{s}$ law at fixed $\alpha$. When the maximum
is at the band centre, $\sup_E\nu(E)$ in
Eq.~\eqref{eq:levySmallEdge} is $\nu(0)$.

 To sum up, the modular energies are confined for every Hamiltonian. At leading order in
$1/m$ the support of $\rst$ lies within $|\beta-\beta_0|\le2\sqrt{2/m}$, Eq.~\eqref{eq:confine}. The
bound is set by the baths alone, because the evolution with $H_0$ can only average the bath noise,
$\|Y\|\le\|X\|$. The exact modular energies obey the realisation-dependent bound
\eqref{eq:exactBetaBound}, since $\beta_j$ is the log-ratio of the loss and gain rates of the $j$-th
natural orbital, Eq.~\eqref{eq:exactRateRatio}. L\'evy Hamiltonians, with spectra growing as
$N^{1/\alpha}$, stay below the same bound [Fig.~\ref{fig:compact}(b)] and are described by the vector
Dyson equation with their exact density of states. Their heavy tails do not push the edge outwards. At small
$s$ it is set by the maximum of $\nu$, Eq.~\eqref{eq:levySmallEdge}, as for a bounded spectrum. The
tails leave two other marks: the slow approach $1-\Psi\propto s^{-\alpha}$ to the semicircle,
Eq.~\eqref{eq:onemPsi}, instead of $s^{-2}$, and a non-analytic centre of $\rst$, a divergent peak for
$\alpha<1$ and a cusp for $1<\alpha<2$, Eq.~\eqref{eq:tail2}.

\section{Number fluctuations, entropy and capacity of entanglement}\label{sec:fluct}
\subsection{Number fluctuations and entropy}
The spread of the modular energies has a direct macroscopic signature. Since $Q\ge0$ and a generic $H_0$ has no eigenvector in the kernel of $Q$, all
eigenvalues of $H$ lie strictly in the lower half-plane (Sec.~\ref{subsec:cloud}), also at $M<N$. Every
mode is then damped, and the steady state is unique and Gaussian~\cite{Prosen2008}. Since the Lindbladian is
invariant under $\hat c\to e^{i\varphi}\hat c$, it has no pairing correlations. In the basis of
natural orbitals it therefore factorises, as anticipated in Sec.~\ref{sec:model}, into independent modes with occupations
$n_j=1/(1+e^{\beta_j})$, and the particle number is a sum of independent Bernoulli variables:
\begin{equation}\label{eq:varN}
\mathrm{Var}\,\hat N=\sum_jn_j(1-n_j)=N\big[\bar n(1-\bar n)-\sigma_n^2\big],
\end{equation}
where $\bar n$ and $\sigma_n^2$ are the mean and the variance of the occupations over the modes. The
variance is thus below its binomial value $N\bar n(1-\bar n)$, that of $N$ modes with equal
occupations, by the relative amount
\begin{equation}\label{eq:deltaN}
\delta_N=1-\frac{\mathrm{Var}\,\hat N}{N\bar n(1-\bar n)}=\frac{\sigma_n^2}{\bar n(1-\bar n)} :
\end{equation}
a spread of the $\beta_j$ appears as sub-binomial number fluctuations. Since
$\beta_j-\beta_0\sim m^{-1/2}$ is small, the occupations are linear in it to leading order, so that
$\sigma_n^2\simeq[n_0(1-n_0)]^2\sigma_\beta^2$ with $n_0(1-n_0)=\gamma\gt/s^2$, Eq.~\eqref{eq:F0}, and
with the width \eqref{eq:width}
\begin{equation}\label{eq:fano}
\delta_N\simeq n_0(1-n_0)\,\sigma_\beta^2=\frac{\gamma\gt}{s^2}\,\frac2m\,\Psi(s).
\end{equation}
In the same way the concavity of the entropy $\mathcal S=\sum_jh(n_j)$,
$h(n)=-n\ln n-(1-n)\ln(1-n)$, gives an entropy below that of the state in which every mode has the
mean occupation,
\begin{equation}\label{eq:entropy}
\frac{\Delta\mathcal S}{N}=\frac{\mathcal S-Nh(\bar n)}{N}\simeq-\frac{\sigma_n^2}{2\bar n(1-\bar n)}
=-\frac12\,\delta_N .
\end{equation}
The origin is the mismatch between the two baths: gain and
loss couple to different random combinations of the modes.
At a fixed mean occupation $\bar n$, the resulting spread
reduces both the number fluctuations and the entropy relative
to the uniform state
$\hat\rho_{\rm unif}\propto e^{-\beta_{\rm eff}\hat N}$,
where $\beta_{\rm eff}=\ln[(1-\bar n)/\bar n]$.
In general $\beta_{\rm eff}$ differs from $\beta_0$ at order
$1/m$. The leading suppression is largest for balanced baths
and $s\gg W$, where $\delta_N=1/(2m)$.

\begin{figure}[t]
\centering
\includegraphics{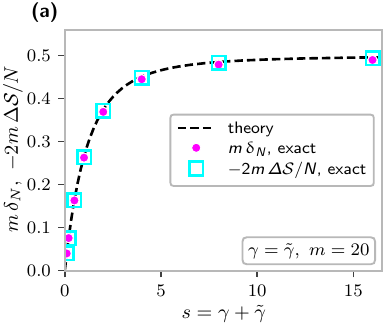}\hspace{1.2em}
\includegraphics{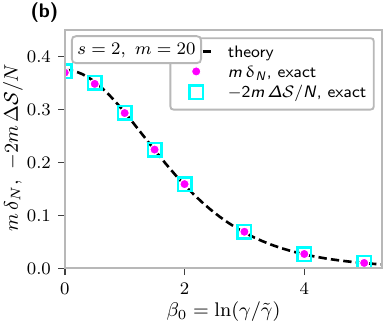}
\caption{Suppression of the particle-number fluctuations, $\delta_N$ (filled symbols), and entropy
deficit, $-2\Delta\mathcal S/N$ (open symbols), both multiplied by $m$, for the GUE $H_0$. Symbols:
exact solutions of \eqref{eq:lyap} ($N=800$, $m=20$, $4$ realisations). Dashed line: Eqs.~\eqref{eq:fano}
and \eqref{eq:entropy} with the analytic $\Psi$, Eq.~\eqref{eq:psi}. (a) Against $s$ for balanced
baths, $\gamma=\gt$. (b) Against $\beta_0=\ln(\gamma/\gt)$ at $s=2$.}
\label{fig:fluct}
\end{figure}

Figure~\ref{fig:fluct} compares \eqref{eq:fano} and \eqref{eq:entropy} with exact solutions of the
Lyapunov equation. Both $\delta_N$ and the entropy deficit $-\Delta\mathcal S$ grow with $s$, following $\Psi(s)$, since a shorter relaxation time leaves
the Hamiltonian less time to average the bath noise. They are largest for balanced baths, where
$n_0(1-n_0)=1/[4\cosh^2(\beta_0/2)]$ is maximal.

\subsection{Capacity of entanglement}\label{subsec:capacity}
The same reasoning applies to any sum over modes. The steady state is $\hat\rho=e^{-\hat K}/Z$
with the modular Hamiltonian $\hat K=\hat c^\dagger\Hst\hat c$, and the variance of $\hat K$,
\begin{equation}
C=\langle\hat K^2\rangle-\langle\hat K\rangle^2,
\end{equation}
is the modular capacity. For a reduced density matrix it is called the capacity of
entanglement~\cite{YaoQi2010}. Here it refers to the full steady state of the open system, and no
bipartition is implied. Below $C$ denotes its average over disorder. It is the heat capacity
$C(T)=\big(\langle\hat K^2\rangle_T-\langle\hat K\rangle_T^2\big)/T^2$ of the family
$\hat\rho_T\propto e^{-\hat K/T}$ at $T=1$, where $T$ is a fictitious temperature. A true heat
capacity cannot be defined here, since neither the reservoirs nor the $\beta_j$ involve a temperature,
and $C$ is its natural substitute. The modes are independent, so
\begin{equation}\label{eq:capacity}
\frac CN=\int\dd\beta\,\rst(\beta)\,c(\beta),\qquad
c(\beta)=\beta^2\,n(\beta)\big[1-n(\beta)\big]=\frac{\beta^2}{4\cosh^2(\beta/2)},
\end{equation}
and expanding about $\beta_0$ with $\sigma_\beta^2=2\Psi(s)/m$ gives

\begin{equation}\label{eq:capexp}
\frac CN\simeq c(\beta_0)+\frac{\Psi(s)}{m}\,c''(\beta_0),\qquad
c''(\beta)=n(1-n)\Big[2-4\beta\tanh\frac\beta2+\beta^2\Big(\frac32\tanh^2\frac\beta2-\frac12\Big)\Big].
\end{equation}
A shift of the mean of the $\beta_j$ would add a term $c'(\beta_0)(\bar\beta-\beta_0)$, but to this order
$\av{\bar\beta}=\beta_0$: the contributions of $F_2$ and of $F_1^2$ to $\av{\tr\Hst}$ at order $1/m$ cancel
exactly, which we also checked numerically. The first term is the value for the state
$e^{-\beta_0\hat N}$. The
correction is a single-mode function $c''(\beta_0)$, which changes sign at $|\beta_0|\approx0.99$ and $|\beta_0|\approx3.79$, times the same factor $\Psi(s)/m$ as in $\delta_N$. For balanced baths, $\beta_0=0$,
$c(0)=0$, and $C/N\simeq\sigma_\beta^2/4=\Psi(s)/2m$ is entirely an effect of the spread of the
$\beta_j$. It then coincides with $\delta_N$ and $-2\Delta\mathcal S/N$, Eqs.~\eqref{eq:fano} and
\eqref{eq:entropy}.

\begin{figure}[t]
\centering
\includegraphics{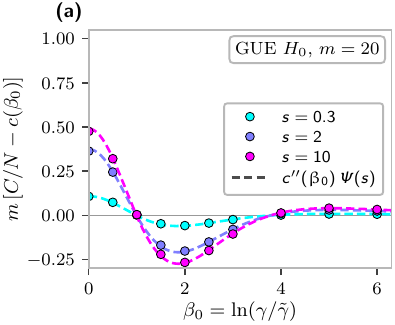}\hspace{1.2em}
\includegraphics{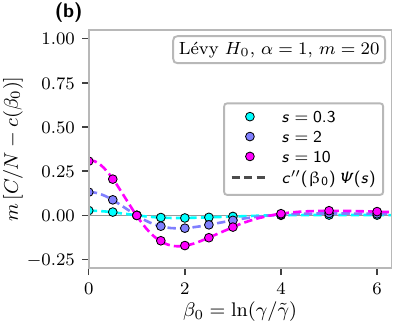}
\caption{Capacity of entanglement in excess of its value $c(\beta_0)$ for the state
$e^{-\beta_0\hat N}$, multiplied by $m$, against $\beta_0=\ln(\gamma/\gt)$ at fixed $s$. Symbols: exact
solutions of \eqref{eq:lyap} ($N=400$, $m=20$, $4$ realisations for the GUE, $8$ for the L\'evy
Hamiltonian with $\alpha=1$). Dashed lines: $c''(\beta_0)\,\Psi(s)$, Eq.~\eqref{eq:capexp}, with the analytic
$\Psi$ for the semicircle and with $\Psi$ from the exact L\'evy density of states \eqref{eq:levydos}.}
\label{fig:capacity}
\end{figure}

Figure~\ref{fig:capacity} shows a comparison of Eq. \eqref{eq:capexp} with exact solutions for the GUE and for a
L\'evy Hamiltonian. The shape of the curves is the same for both, being fixed by the statistics of a
single fermionic mode. The Hamiltonian enters only through the amplitude $\Psi(s)$, which is smaller
for the L\'evy matrix because its heavy tails place fewer neighbours within the window $s$.

\section{Relation to earlier work}\label{sec:earlier}
Random Lindbladians have been studied intensively~\cite{Denisov2019,Can2019,Sa2020,Costa2023},
mostly through the spectrum of the Liouvillian and the purity and level statistics of the steady
state. The distribution of the modular energies has been addressed in two works.

Thompson and Kamenev~\cite{Thompson2023} introduced the model studied here, a random Hamiltonian
with random gain and loss, and formulated its Keldysh field theory. For pure loss they obtained the
spectrum of the decay rates, with the Marchenko--Pastur edges. They had no analytic solution for $\rst$ at
large $N$ and large $m$, and for many channels described it as a Wigner semicircle of width $\propto1/m$,
centred at a position set by $\gt/\gamma$. Our results correct this picture. The centre is indeed
$\beta_0=\ln(\gamma/\gt)$, Eq.~\eqref{eq:F0}, but the radius of the semicircle is $2\sqrt{2/m}$,
Eq.~\eqref{eq:larges}, so that $\sigma_\beta^2$, not the width, is $\propto1/m$. Moreover, $\rst$ is a
semicircle only for $s\gg W$. At smaller $s$ it depends on the density of states of $H_0$ through the
vector Dyson equation and tends to the law \eqref{eq:smalls} for $s\to0$.

Costa \emph{et al.}~\cite{Costa2023} study random quadratic Liouvillians with a Hamiltonian of class D
and a single set of $M$ linear jump operators. Their effective Hamiltonian $\Omega$, defined by
$\hat\rho\propto e^{\hat\Omega}$, plays the role of our $-\Hst$. Without separate gain and loss there is
no centre $\beta_0$, and the spectrum of $\Omega$ is symmetric about zero. At weak dissipation they
find a Gaussian of width $\sqrt{2/M}$ from a central limit theorem. This is the regime of a
dissipation weaker than the level spacing, $s\ll\Delta$, in which every eigenstate of $H_0$ is a
natural orbital and its modular energy is set by its own couplings to the baths. Our results hold in
the opposite regime $\Delta\ll s$, where levels within $s$ of each other mix and the spread is set by
$\Psi(s)$. At strong dissipation Costa \emph{et al.} express the steady state through the eigenvalues
and eigenvectors of a Wishart matrix but do not obtain its distribution explicitly. The corresponding
limit here, $s\gg W$, gives the semicircle \eqref{eq:larges} at large $m$.

For L\'evy Hamiltonians we use the density of states of
Refs.~\cite{Cizeau1994,BenArous2008,Safonova2025}. We use the result of the supersymmetric calculation of Ref.~\cite{Safonova2025}. The cavity
approach of Ref.~\cite{Cizeau1994} reproduces it once the asymmetry of the stable law of the cavity
self-energy is weighted with $|G|^{\alpha/2}$ (Sec.~\ref{subsec:levy}). The unweighted asymmetry used
in Ref.~\cite{Cizeau1994} changes the density of states in the bulk.

 These works study a closed L\'evy matrix: its density of states and the localisation of
its eigenstates. Here the L\'evy matrix is the Hamiltonian of an open system, and its eigenstates play
no role, since $\rst$ depends on $H_0$ only through the density of states $\nu$. In this role the unbounded spectrum does
not widen the distribution of the modular energies. It stays within the bound \eqref{eq:confine} and
is even narrower than for the GUE, because most levels of a broad spectrum have few neighbours within
$s$. The heavy tails show up only at the centre of $\rst$ and in the slow approach $1-\Psi\propto
s^{-\alpha}$ to the semicircle (Sec.~\ref{sec:confine}).

The universality found here differs from that of closed chaotic
systems~\cite{Beenakker1997,Alhassid2000}, where the spectral statistics depend on the symmetry class
and not on microscopic details. Here the level statistics of $H_0$, and with them its symmetry class,
drop out, and $\rst$ depends only on the density of states, i.e.\ on precisely the non-universal part
of the spectrum of a closed system.

\section{Conclusion}\label{sec:conclusion}
The distribution of the modular energies $\beta_j=\ln[(1-n_j)/n_j]$, the log-odds of the occupations
of the natural orbitals, in the steady state of
the Thompson--Kamenev model is described at large channel number $m$ by a vector Dyson equation, with closed-form solutions in limiting regimes. The Lyapunov
equation can be solved as a series in $m^{-1/2}$. At leading order $\Hst=\beta_0\id+Y$, where the
centre $\beta_0=\ln(\gamma/\gt)$ depends only on the ratio of the rates, and $Y$ is the bath noise,
evolved by the Hamiltonian and averaged over the relaxation time $1/s$, $s=\gamma+\gt$.

The variance of the distribution is set by the Laplace transform, at the rate $s$, of the squared
return amplitude of the Hamiltonian, Eq.~\eqref{eq:width}. It depends on $H_0$ only through its density of
states $\nu$, and for the GUE it is given by complete elliptic integrals. The shape follows from the
vector Dyson equation of a Wigner-type matrix with a Lorentzian variance profile of width $s$. It
again contains $H_0$ only through $\nu$: chaotic and integrable Hamiltonians with the same density of
states give the same $\rst$. For the GUE the shape interpolates between a semicircle of radius
$2\sqrt{2/m}$ for $s\gg W$ and a distinct law, Eq. \eqref{eq:smalls}, for $s\ll W$, expressed through complete elliptic
integrals and built as a superposition of local semicircles across the band.

The edge of the distribution is a square root at every $s>0$, Eq.~\eqref{eq:root}, with a strictly
positive amplitude. For $s\gg W$ the first correction rescales the distribution without deforming it.
For $s\ll W$ the square root retreats into a boundary layer of width $\sqrt s$ at the band centre,
governed by the parameter-free equation \eqref{eq:universal}, and the linear edge of the $s\to0$ law
is never exact at finite $s$: the density is linear only beyond a distance of order $s\,\sigma_\beta$ from
the edge and reaches the $s\to0$ law for $R_0-b\gg sR_0\ln(1/s)$. Both scales vanish as $s\to0$.

The leading large-$m$ modular density has support within
$|\beta-\beta_0|\le2\sqrt{2/m}$ for any Hamiltonian. The exact
contraction $\|Y\|\le\|X\|$ explains why Hamiltonian evolution
cannot increase the norm of the linearised bath noise.
For the exact modular energies, a separate realisation-dependent
bound follows from the extremal eigenvalues of the dissipators,
Eq.~\eqref{eq:exactBetaBound}. L\'evy matrices, with unbounded spectra and
eigenstates localised in the spectral tails, confirm this and are described by the same theory once their exact density of
states is used. Their heavy tails leave a mark at the centre of the distribution, a divergent peak
for $\alpha<1$ and a cusp for $1<\alpha<2$, and in a slow, $s^{-\alpha}$ approach to the semicircle.

Finally, the spread of the $\beta_j$ has direct signatures in quantities that are sums over modes:
the number fluctuations are sub-binomial, Eq.~\eqref{eq:fano}, the entropy is lower by
$\frac12\delta_N$ per mode than that of a state with uniform occupations, and the capacity of
entanglement acquires a correction $c''(\beta_0)\Psi(s)/m$ whose shape is universal.

These results amend the large-$m$ statement of Ref.~\cite{Thompson2023}: the centre is indeed
$\ln(\gamma/\gt)$, but the width scales as $m^{-1/2}$ rather than $m^{-1}$, and the shape is a
semicircle only for $s\gg W$.

\section*{Acknowledgements}
We thank A.~Kamenev for fruitful discussions and M.~Shishkin for helpful ideas on the derivation
of the Keldysh $\sigma$-model. M. V. P. acknowledges the hospitality during the ``Quantum matter beyond ergodicity and thermalization'' where a part of this work has been performed. 
The authors acknowledge personal support from the Foundation for the Advancement of Theoretical Physics and Mathematics ``BASIS''. During the preparation of this work, the authors used Claude (Anthropic) for language editing and assistance with code development. The authors reviewed and edited all output and take full responsibility for the content of the publication.


\paragraph{Funding information}

This work was partially supported  by  the Russian Science Foundation under Grant No. \href{https://rscf.ru/project/26-12-00052/}{26-12-00052} (computations of the average density) as well as by Basic Research Program of HSE under Grant No. HSE-BR-2025-57 (computation of the boundary for eigenvalues of non-Hermitian Hamiltonian). 


\begin{appendix}
\numberwithin{equation}{section}

\section{Wishart correlators}\label{app:wishart}
The disorder weight
$P\propto\exp[-\frac N2\tr H_0^2-\frac M\gamma\sum|\mu_{vj}|^2-\frac M\gt\sum|\tilde\mu_{vj}|^2]$
is Gaussian, U(1)-invariant and factorised. The only non-vanishing contractions of the bath amplitudes are
$\langle\mu^*_{vj}\mu_{wk}\rangle=(\gamma/M)\delta_{vw}\delta_{jk}$ and the analogous one for
$\tilde\mu$.

\paragraph{Two points.} The second moment is $\left\langle \Gamma_{jk} \Gamma_{lm}\right\rangle = \sum_{v,w}  \left\langle\mu^*_{v j} \mu_{vk}\mu^*_{w l} \mu_{w m} \right\rangle$. Wick's theorem with the contraction $\langle\mu^*_{vj}\mu_{wk}\rangle$ gives two pairings, which with the
flavour deltas kept visible read
\begin{equation}\label{eq:pairings}
\begin{aligned}
\text{(P1)}\ \ \big\langle\mu^*_{vj}\mu_{vk}\big\rangle\big\langle\mu^*_{wl}\mu_{wm}\big\rangle
&=\Big(\frac{\gamma}{M}\Big)^{\!2}\,\delta_{vv}\delta_{jk}\;\delta_{ww}\delta_{lm}
&&\xrightarrow{\ \textstyle\sum_{v,w}\ }\
M^2\Big(\frac{\gamma}{M}\Big)^{\!2}\delta_{jk}\delta_{lm}=\gamma^2\,\delta_{jk}\delta_{lm},\\[2pt]
\text{(P2)}\ \ \big\langle\mu^*_{vj}\mu_{wm}\big\rangle\big\langle\mu^*_{wl}\mu_{vk}\big\rangle
&=\Big(\frac{\gamma}{M}\Big)^{\!2}\,\delta_{vw}\delta_{jm}\;\delta_{wv}\delta_{lk}
&&\xrightarrow{\ \textstyle\sum_{v,w}\ }\
M\Big(\frac{\gamma}{M}\Big)^{\!2}\delta_{jm}\delta_{kl}=\frac{\gamma^2}{M}\,\delta_{jm}\delta_{kl}.
\end{aligned}
\end{equation} 
Therefore the second moment is
\begin{equation}
    \left\langle \Gamma_{jk} \Gamma_{lm}\right\rangle = \gamma^2\,\delta_{jk}\delta_{lm} + \frac{\gamma^2}{M}\,\delta_{jm}\delta_{kl} \quad \Rightarrow \quad  \left\langle \xi_{jk} \xi_{lm}\right\rangle = \frac{1}{M} \delta_{jm} \delta_{kl}.
\end{equation}
The same expression with $\gamma\to\gt$ holds for $\Gt$, so $\xit$ has the same correlator.

Because the weight factorises, $\av{\xi\,\xit}=0$, and all mixed
cumulants vanish. Therefore
$\av{X_{jk}X_{lm}}=(2/M)\delta_{jm}\delta_{kl}$.

\paragraph{Higher cumulants.} Wick's theorem gives the higher moments in the same way. For $n=3$ the
connected pairings are the two cyclic ones,
\begin{equation}\label{eq:3pt}
\big\langle\xi_{j_1k_1}\xi_{j_2k_2}\xi_{j_3k_3}\big\rangle
=\frac{1}{M^2}\Big[\delta_{k_1j_2}\delta_{k_2j_3}\delta_{k_3j_1}
+\delta_{k_1j_3}\delta_{k_3j_2}\delta_{k_2j_1}\Big].
\end{equation}
In general the $n$-th cumulant is a sum of such cyclic products of Kronecker deltas with the
prefactor $1/M^{n-1}$: each pairing carries $(\gamma/M)^n$, and the sum over the channel gives one
factor $M$. For $X=\xi-\xit$ the odd cumulants cancel, and the first non-Gaussian one is the fourth.
Inserted in $\av{\frac1N\tr X^n}$, a cyclic product leaves at most $n$ free indices, so the $n$-th
cumulant contributes at most $N^{n-1}/M^{n-1}=m^{1-n}$. Against the scale $\sigma_\beta^n\sim m^{-n/2}$ set
by the width this is $m^{1-n/2}$, i.e.\ $1/m$ for the fourth cumulant, which is why the expansion
parameter is $1/m$ and not $1/M$. The cyclic delta structure is unitarily invariant, by the argument
of property (c) of Sec.~\ref{sec:model}, so the same holds for $X'$ in the eigenbasis of $H_0$.
The factors $r_{jk}$ of $Y_{jk}=r_{jk}X'_{jk}$ do not change the index structure and have $|r_{jk}|\le1$.
The non-Gaussian cumulants therefore correct $\rst$ only at relative order $1/m$. This is smaller than
the terms of absolute order $1/m$ in $\Hst$, i.e.\ of relative order $m^{-1/2}$, neglected in
\eqref{eq:master}.

\section{Saddle point of the Keldysh action and the boundary of the spectrum of \texorpdfstring{$H$}{H}}\label{app:sigma}
In this appendix we derive the boundary \eqref{eq:cloudbdy} of the spectrum of $H=H_0-iQ$. We use
the Keldysh action of the Lindbladian dynamics~\cite{Thompson2023,Kamenev2011}, average it over
$H_0$ and over the baths, decouple the resulting quartic terms by a matrix field, and analyse the
saddle point at large $N$. We do not derive the nonlinear $\sigma$-model, i.e.\ the theory of
the fluctuations about the saddle point. At $N\to\infty$ the averaged resolvent and the boundary of the
cloud are fixed by the saddle-point equation alone. The GUE $H_0$ is normalised as in Eq.~\eqref{eq:nusc},
$\langle(H_0)_{ij}(H_0)_{kl}\rangle=\delta_{il}\delta_{jk}/N$, and the jump amplitudes are as in
Sec.~\ref{sec:model}, $\langle\mu^*_{vj}\mu_{wk}\rangle=(\gamma/M)\delta_{vw}\delta_{jk}$ and the
same with $\gamma\to\gt$ for $\tilde\mu$. We use $g=\sqrt{\gamma/m}$, $\tilde g=\sqrt{\gt/m}$, the
index $\sigma=\pm$ with $\gamma_+=\gamma$, $\gamma_-=\gt$, and the Fourier convention
$\psi(t)=\int\frac{\dd\epsilon}{2\pi}\,e^{-i\epsilon t}\psi(\epsilon)$,
$\bar\psi(t)=\int\frac{\dd\epsilon}{2\pi}\,e^{i\epsilon t}\bar\psi(\epsilon)$. Below, $\int_\epsilon$
stands for $\int\dd\epsilon/2\pi$.

In the Larkin--Ovchinnikov basis, $\psi=(\psi^1,\psi^2)$ with $\psi^{1,2}=(\psi^+\pm\psi^-)/\sqrt2$ and
$\bar\psi^{1,2}=(\bar\psi^+\mp\bar\psi^-)/\sqrt2$, the Keldysh action of the model of
Sec.~\ref{sec:model} reads~\cite{Thompson2023}
\begin{equation}\label{eq:Skeld}
S=\int_\epsilon\bar\psi(\epsilon)\big(\epsilon\,\sigma_0-\sigma_0H_0+i\sigma_3Q+i\sigma_+D\big)\psi(\epsilon),
\qquad Z=\int\mathrm D\bar\psi\,\mathrm D\psi\;e^{iS}=1,
\end{equation}
where $\sigma_i$ are the Pauli matrices in Keldysh space and $\sigma_+=(\sigma_1+i\sigma_2)/2$. The
retarded block of the action contains $\epsilon-H_0+iQ=\epsilon-H$. With Eq.~\eqref{eq:QD}, the bath
part splits into a loss and a gain term,
\begin{equation}
\begin{gathered}
i\sigma_3Q+i\sigma_+D=i\,\Gamma\,\mathcal O_++i\,\Gt\,\mathcal O_-,\\
\mathcal O_\pm=\tfrac12\sigma_3\pm\sigma_+=U_\pm\Delta U_\pm^{-1},\qquad
U_\pm=\begin{pmatrix}1&\mp1\\0&1\end{pmatrix},\qquad \Delta=\mathrm{diag}\big(\tfrac12,-\tfrac12\big).
\end{gathered}
\end{equation}
The average over the GUE gives
\begin{equation}
\big\langle e^{-i\int_\epsilon\bar\psi H_0\psi}\big\rangle
=\exp\Big(-\frac1{2N}\iint_{\epsilon,\epsilon_1}
\bar\psi_i(\epsilon)\psi_j(\epsilon)\bar\psi_j(\epsilon_1)\psi_i(\epsilon_1)\Big)
=\exp\Big(\frac1{2N}\,\mathrm{Tr}\,\Lambda^2\Big),
\end{equation}
where $\Lambda_{\epsilon\epsilon_1}=\sum_j\psi_j(\epsilon)\bar\psi_j(\epsilon_1)$ is a $2\times2$
matrix in Keldysh space, and $\mathrm{Tr}$ runs over Keldysh indices and energies. For the baths,
the loss term is a sum over channels of products of two Keldysh vectors,
\begin{equation}
\bar\psi_i\Gamma_{ij}\mathcal O_+\psi_j=\sum_v\bar\Psi_v\Psi_v ,\qquad
\Psi_v=\Delta^{1/2}U_+^{-1}\mu_{vj}\psi_j,\qquad \bar\Psi_v=\bar\psi_i\mu^*_{vi}\,U_+\Delta^{1/2},
\end{equation}
and we decouple it by an auxiliary Grassmann field $\chi_v$,
\begin{equation}
\exp\Big(-\int_\epsilon\bar\Psi_v\Psi_v\Big)=\int\mathrm D\bar\chi\,\mathrm D\chi\,
\exp\Big(\int_\epsilon\big[\bar\chi_v\chi_v-\bar\chi_v\Psi_v-\bar\Psi_v\chi_v\big]\Big),
\end{equation}
and the gain term in the same way, by a field $\tilde\chi_v$ and with $U_+\to U_-$. The action is now
linear in $\mu$ and $\tilde\mu$, and the Gaussian average over them gives
\begin{equation}
iS_\mu=\frac{\gamma}{2M}\iint_{\epsilon,\epsilon_1}\Big(
\bar\chi_v\Delta^{1/2}U_+^{-1}\psi_j\cdot\bar\psi_jU_+\Delta^{1/2}\chi_v
+\bar\psi_jU_+\Delta^{1/2}\chi_v\cdot\bar\chi_v\Delta^{1/2}U_+^{-1}\psi_j\Big),
\end{equation}
and $iS_{\tilde\mu}$ with $\gamma\to\gt$, $\chi\to\tilde\chi$, $U_+\to U_-$. In terms of the $2\times2$
fields
\begin{equation}
P_{\epsilon\epsilon_1}=U_+\Delta^{1/2}\chi_v(\epsilon)\bar\chi_v(\epsilon_1)\Delta^{1/2}U_+^{-1},\qquad
\tilde P_{\epsilon\epsilon_1}=U_-\Delta^{1/2}\tilde\chi_v(\epsilon)\bar{\tilde\chi}_v(\epsilon_1)\Delta^{1/2}U_-^{-1},
\end{equation}
all the quartic terms combine into
\begin{equation}
\begin{aligned}
iS_{\rm int}&=\frac1{2N}\iint_{\epsilon,\epsilon_1}\Big(\tr\Lambda_{\epsilon\epsilon_1}\Lambda_{\epsilon_1\epsilon}
-\frac\gamma m\tr\big[\Lambda_{\epsilon\epsilon_1}P_{\epsilon_1\epsilon}+P_{\epsilon\epsilon_1}\Lambda_{\epsilon_1\epsilon}\big]
-\frac\gt m\tr\big[\Lambda_{\epsilon\epsilon_1}\tilde P_{\epsilon_1\epsilon}+\tilde P_{\epsilon\epsilon_1}\Lambda_{\epsilon_1\epsilon}\big]\Big)\\
&=\frac1{2N}\iint_{\epsilon,\epsilon_1}\tr\,\mathcal M_{\epsilon\epsilon_1}\mathcal M_{\epsilon_1\epsilon},
\end{aligned}
\end{equation}
with the $6\times6$ matrix (Keldysh $\otimes$ species $\psi,\chi,\tilde\chi$)
\begin{equation}
\mathcal M_{\epsilon\epsilon_1}=\begin{pmatrix}
\Lambda_{\epsilon\epsilon_1}&g\,\Lambda_{\epsilon\epsilon_1}&\tilde g\,\Lambda_{\epsilon\epsilon_1}\\
-g\,P_{\epsilon\epsilon_1}&0&0\\
-\tilde g\,\tilde P_{\epsilon\epsilon_1}&0&0\end{pmatrix}.
\end{equation}

We decouple $iS_{\rm int}$ by a matrix field $\mathcal Q$ in the same $6\times6$ space,
\begin{equation}
e^{iS_{\rm int}}=\int\mathrm D\mathcal Q\,\exp\Big(-\frac N2\iint_{\epsilon,\epsilon_1}
\tr\,\mathcal Q_{\epsilon\epsilon_1}\mathcal Q_{\epsilon_1\epsilon}
+\iint_{\epsilon,\epsilon_1}\tr\,\mathcal Q_{\epsilon\epsilon_1}\mathcal M_{\epsilon_1\epsilon}\Big),
\end{equation}
which follows by completing the square under the trace and the energy integrals. Only the blocks
of $\mathcal Q$ that couple to non-zero blocks of $\mathcal M$ matter, namely $\mathcal Q^{(11)}$,
$\mathcal Q^{(12)}$, $\mathcal Q^{(13)}$, $\mathcal Q^{(21)}$ and $\mathcal Q^{(31)}$. The others
appear only in $\tr\mathcal Q^2$ and decouple. The fermions now enter quadratically,
\begin{equation}
\begin{aligned}
iS_\psi&=i\iint_{\epsilon,\epsilon_1}\bar\psi(\epsilon)\Big(\epsilon\,\sigma_0\,2\pi\delta(\epsilon-\epsilon_1)
+i\big[\mathcal Q^{(11)}+g\,\mathcal Q^{(21)}+\tilde g\,\mathcal Q^{(31)}\big]_{\epsilon\epsilon_1}\Big)\psi(\epsilon_1),\\
iS_\chi&=\iint_{\epsilon,\epsilon_1}\bar\chi(\epsilon)\Big(\sigma_0\,2\pi\delta(\epsilon-\epsilon_1)
+g\,\Delta^{1/2}U_+^{-1}\mathcal Q^{(12)}_{\epsilon\epsilon_1}U_+\Delta^{1/2}\Big)\chi(\epsilon_1),
\end{aligned}
\end{equation}
and $iS_{\tilde\chi}$ with $g\to\tilde g$, $U_+\to U_-$ and $\mathcal Q^{(12)}\to\mathcal Q^{(13)}$.
Integrating out $\psi$ ($N$ flavours) and $\chi_v,\tilde\chi_v$ ($M=mN$ flavours each) gives a theory
of $\mathcal Q$ alone, $Z=\int\mathrm D\mathcal Q\,e^{iS_{\rm eff}[\mathcal Q]}$, with
\begin{equation}\label{eq:Seff}
iS_{\rm eff}=N\Big[-\frac12\,\mathrm{Tr}\,\mathcal Q^2+\mathrm{Tr}\ln\big(\mathcal G^\psi\big)^{-1}
+m\,\mathrm{Tr}\ln\big(\mathcal G^\chi\big)^{-1}+m\,\mathrm{Tr}\ln\big(\mathcal G^{\tilde\chi}\big)^{-1}\Big],
\end{equation}
\begin{equation}
\begin{gathered}
\big(\mathcal G^\psi\big)^{-1}=\epsilon\,\sigma_0+i\big(\mathcal Q^{(11)}+g\,\mathcal Q^{(21)}+\tilde g\,\mathcal Q^{(31)}\big),\\
\big(\mathcal G^\chi\big)^{-1}=\sigma_0+g\,\Delta^{1/2}U_+^{-1}\mathcal Q^{(12)}U_+\Delta^{1/2},\qquad
\big(\mathcal G^{\tilde\chi}\big)^{-1}=\sigma_0+\tilde g\,\Delta^{1/2}U_-^{-1}\mathcal Q^{(13)}U_-\Delta^{1/2}.
\end{gathered}
\end{equation}
The large prefactor $N$ justifies the saddle-point approximation. The causal structure of the theory
is fixed by the infinitesimal $\epsilon\sigma_0\to\epsilon\sigma_0+i0\,\sigma_3$, and below it is replaced
by a finite term.

Varying \eqref{eq:Seff} with $\delta\,\mathrm{Tr}\ln A=\mathrm{Tr}\,A^{-1}\delta A$ gives
\begin{equation}
\begin{gathered}
\mathcal Q^{(11)}=i\mathcal G^\psi,\qquad \mathcal Q^{(12)}=ig\,\mathcal G^\psi,\qquad
\mathcal Q^{(13)}=i\tilde g\,\mathcal G^\psi,\\
\mathcal Q^{(21)}=mg\,U_+\Delta^{1/2}\mathcal G^\chi\Delta^{1/2}U_+^{-1},\qquad
\mathcal Q^{(31)}=m\tilde g\,U_-\Delta^{1/2}\mathcal G^{\tilde\chi}\Delta^{1/2}U_-^{-1}.
\end{gathered}
\end{equation}
For a saddle point diagonal in energy, $\mathcal G=\mathcal G^\psi_{\epsilon}$ obeys, after
eliminating the auxiliary fields with $U_\pm\Delta^{-1}U_\pm^{-1}=4\mathcal O_\pm$,
\begin{equation}\label{eq:SCBA}
\mathcal G=\Big[\epsilon\,\sigma_0-\mathcal G+i\sum_{\sigma=\pm}\gamma_\sigma\,\mathcal V_\sigma^{-1}\Big]^{-1},
\qquad \mathcal V_\sigma=4\mathcal O_\sigma+\frac{i\gamma_\sigma}{m}\,\mathcal G .
\end{equation}
This saddle-point equation is the self-consistent Born approximation for our model. The term $-\mathcal G$ is the
rainbow diagram of the GUE, and the terms with $\mathcal V_\sigma$ resum the ladders of the Wishart
couplings.

Equation~\eqref{eq:SCBA} respects the upper-triangular (causal) Keldysh structure,
$\mathcal G=\big(\begin{smallmatrix}\mathcal G^{\rm R}&\mathcal G^{\rm K}\\0&\mathcal G^{\rm A}\end{smallmatrix}\big)$,
and then
\begin{equation}
\mathcal V_\sigma=\begin{pmatrix}2\alpha_\sigma&\beta_\sigma\\0&-2\delta_\sigma\end{pmatrix},\qquad
\alpha_\sigma=1+\frac{i\gamma_\sigma}{2m}\mathcal G^{\rm R},\quad
\delta_\sigma=1-\frac{i\gamma_\sigma}{2m}\mathcal G^{\rm A},\quad
\beta_\sigma=4\sigma+\frac{i\gamma_\sigma}{m}\mathcal G^{\rm K}.
\end{equation}
Its retarded component is Eq.~\eqref{eq:Gcloud} with $\mathcal G^{\rm R}=G$,
\begin{equation}\label{eq:causal}
\frac1{\mathcal G^{\rm R}}=\epsilon-\mathcal G^{\rm R}+\sum_{\sigma=\pm}\frac{i\gamma_\sigma}{2\alpha_\sigma},
\end{equation}
and the advanced one is its complex conjugate. Continued to complex $\epsilon=z$, the holomorphic
solution of \eqref{eq:causal} is the averaged resolvent outside the cloud, and the boundary is where a
non-causal solution branches off it.

\textit{Boundary.} To move the energy into the lower half-plane we add to the action a uniform
damping term, the one that $N$ channels $\hat L_{1,M+j}=\sqrt\varkappa\,\hat c_j$,
$\hat L_{2,M+j}=\sqrt\varkappa\,\hat c_j^\dagger$ with $\varkappa>0$ would produce. It shifts $Q\to Q+\varkappa\id$, i.e.\
$\epsilon\sigma_0\to\epsilon\sigma_0+i\varkappa\sigma_3$. We then continue the coefficient analytically to $\varkappa<0$.
This is a formal source for the resolvent, not a physical GKSL generator.
Treating $z=\epsilon+i\varkappa$ and $\bar z=\epsilon-i\varkappa$ as independent and coupling the two
blocks by a regulator $\eta$, we have a replacement in Eq. \eqref{eq:SCBA}
\begin{equation}
\epsilon\,\sigma_0\;\longrightarrow\;\begin{pmatrix}z&i\eta\\ i\eta&\bar z\end{pmatrix}
\end{equation}
and allow $\mathcal G_{21}\neq0$. Since
$\det\mathcal V_\sigma=-4\alpha_\sigma\delta_\sigma-(i\gamma_\sigma/m)\,\beta_\sigma\,\mathcal G_{21}$ is
linear in $\mathcal G_{21}$, the $(2,1)$ component of Eq. \eqref{eq:SCBA}, linearised about the causal
solution, closes on itself:
\begin{equation}\label{eq:B0}
\mathcal G_{21}\,\mathcal B_0=i\eta,\qquad
\mathcal B_0=1-\frac1{\mathcal G^{\rm R}(z)\,\mathcal G^{\rm A}(\bar z)}
+\sum_{\sigma=\pm}\frac{\gamma_\sigma^2}{4m\,\alpha_\sigma\delta_\sigma}.
\end{equation}
If $\mathcal B_0\neq0$, $\mathcal G_{21}\to0$ as $\eta\to0$ and the causal solution holds: $z$
lies outside the cloud. A non-causal solution with $\mathcal G_{21}\neq0$ appears for
$\mathcal B_0=0$. With $\mathcal G^{\rm A}(\bar z)=\overline{\mathcal G^{\rm R}(z)}$ and
$\delta_\sigma=\bar\alpha_\sigma$ this condition is real,
\begin{equation}
\frac1{|\mathcal G^{\rm R}(z)|^2}=1+\sum_{\sigma=\pm}\frac{\gamma_\sigma^2}{4m\,|\alpha_\sigma|^2},
\end{equation}
which is the boundary \eqref{eq:cloudbdy}.

\textit{Large $m$: the ellipse.} At $m\gg1$, to first order in $1/m$,
$i\gamma_\sigma/2\alpha_\sigma\simeq i\gamma_\sigma/2+\gamma_\sigma^2G/4m$ and
$|\alpha_\sigma|^2\simeq1$. With $v=(\gamma^2+\gt^2)/4m$, Eq.~\eqref{eq:causal} becomes
$1/G=z+is/2-(1-v)G$ and the boundary becomes $|G|^2=1/(1+v)$. Setting
$G=e^{i\theta}/\sqrt{1+v}$,
\begin{equation}
\mathrm{Re}\,z=\frac{2\cos\theta}{\sqrt{1+v}},\qquad
\mathrm{Im}\,z+\frac s2=-\frac{2v\sin\theta}{\sqrt{1+v}},
\end{equation}
which is the ellipse \eqref{eq:ellipse}.

\textit{Pure loss.} Without the Hamiltonian and the gain bath, $H=-i\Gamma/2$, the terms $-G$ and $1$
drop out, and with $u=i\gamma G/2m$ the equations read
$z=(i\gamma/2)\big[1/(mu)-1/(1+u)\big]$ and $|1+u|^2=m|u|^2$. On the imaginary axis, $z=-iq$, $u$ is
real, so $1+u=\pm\sqrt m\,u$, and
\begin{equation}
q_\pm=\frac{\gamma}{2m}\big(1\pm\sqrt m\big)^2 .
\end{equation}
These are the edges of the Marchenko--Pastur law of $Q=\Gamma/2$~\cite{Marchenko1967}, first obtained for this model by Thompson and Kamenev~\cite{Thompson2023}.

\section{Derivation of the vector Dyson equation}\label{app:cavity}
In this appendix we derive Eqs.~\eqref{eq:gj} and \eqref{eq:dyson} from the cavity equation
\eqref{eq:cavity}. As in Sec.~\ref{sec:dyson}, $H_0$ is held fixed, $\langle\cdots\rangle$ denotes
the average over the baths, and $Y$ is the Wigner-type matrix with the covariance \eqref{eq:Ycov}.

We start from the relation between the resolvent and the density. For $\lambda=b+i\eta$ with $\eta>0$
the resolvent is $G(\lambda)=(\lambda-Y)^{-1}$. In the eigenbasis of $Y$, $Y\psi_\alpha=b_\alpha\psi_\alpha$, the resolvent is
$G_{jj}=\sum_\alpha|\psi_\alpha(j)|^2/(\lambda-b_\alpha)$, and therefore
\begin{equation}\label{eq:resdens}
-\frac1\pi\,\mathrm{Im}\,\frac1N\tr G(b+i\eta)=\frac1N\sum_\alpha\frac{\eta/\pi}{(b-b_\alpha)^2+\eta^2}
\;\xrightarrow{\ \eta\to0^+\ }\;\frac1N\sum_\alpha\delta(b-b_\alpha),
\end{equation}
which is the spectral density of $Y$, i.e.\ $\rst$ in the variable $b=\beta-\beta_0$. It is thus
sufficient to find the diagonal elements of $G$. To control the errors below we need two
properties of the resolvent. Since $G-G^\dagger=(\bar\lambda-\lambda)GG^\dagger$, the matrix
$\mathrm{Im}\,G=(G-G^\dagger)/2i=-\eta\,GG^\dagger$ is negative semidefinite, and its diagonal
element gives the Ward identity
\begin{equation}\label{eq:ward}
\sum_l|G_{kl}|^2=(GG^\dagger)_{kk}=\frac{|\mathrm{Im}\,G_{kk}|}{\eta}.
\end{equation}

To obtain an equation for $G_{jj}$, we single out the index $j$ and write $\lambda-Y$ in block form,
\begin{equation}
\lambda-Y=\begin{pmatrix}\lambda-Y_{jj} & -y_j^\dagger\\ -y_j & \lambda-Y^{(j)}\end{pmatrix},
\end{equation}
where $y_j=(Y_{kj})_{k\ne j}$ is the $j$-th column without its diagonal element and $Y^{(j)}$ is the
$(N-1)\times(N-1)$ matrix with row and column $j$ removed. The Schur complement formula for the
upper-left element of the inverse of a block matrix then yields Eq.~\eqref{eq:cavity}, with
$\Sigma_j=y_j^\dagger G^{(j)}y_j$ and $G^{(j)}=\big(\lambda-Y^{(j)}\big)^{-1}$. This exact
relation is useful because $G^{(j)}$ is a function of $Y^{(j)}$ only, whereas $y_j$ consists of the
elements of column $j$. Since the entries of different pairs are independent, $y_j$ is
statistically independent of $G^{(j)}$, and we may average over $y_j$ at fixed $G^{(j)}$. Before
doing so, we note two simple facts. The diagonal element has variance $\av{Y_{jj}^2}=2/M$, hence
$Y_{jj}=O\big((mN)^{-1/2}\big)$ is negligible against $\lambda\sim\sigma_\beta$. Moreover,
$\mathrm{Im}\,\Sigma_j=y_j^\dagger\,(\mathrm{Im}\,G^{(j)})\,y_j\le0$, so that the denominator in
\eqref{eq:cavity} always has imaginary part $\ge\eta$ and never comes close to zero.

We now average $\Sigma_j$ over $y_j$ at fixed $G^{(j)}$. By \eqref{eq:Ycov},
$\av{Y_{jk}Y_{lj}}=\delta_{kl}\av{|Y_{jk}|^2}$, and there is no anomalous contraction
$\av{Y_{jk}Y_{jl}}$. Consequently, only the diagonal terms $k=l$ survive, and
\begin{equation}\label{eq:Qmean}
\av{\Sigma_j}_{y_j}=\frac1N\sum_{k\ne j}S(E_j,E_k)\,G^{(j)}_{kk}.
\end{equation}
Although $\Sigma_j$ contains $N^2$ terms, it is close to this mean, i.e.\ it self-averages. To
see this, we compute its variance. For Gaussian $y_j$ with independent components of variances
$S(E_j,E_k)/N\le 2/(mN)$ we find
\begin{equation}\label{eq:Qvar}
\begin{aligned}
\mathrm{Var}_{y_j}\Sigma_j&=\frac1{N^2}\sum_{k,l\ne j}S(E_j,E_k)\,S(E_j,E_l)\,\big|G^{(j)}_{kl}\big|^2\\
&\le\Big(\frac2{mN}\Big)^2\sum_{k,l}\big|G^{(j)}_{kl}\big|^2
=\frac{4}{m^2}\,\frac{1}{N\eta}\,\frac1N\sum_k\big|\mathrm{Im}\,G^{(j)}_{kk}\big|,
\end{aligned}
\end{equation}
where in the last step we used the Ward identity \eqref{eq:ward}. At fixed $\eta>0$, the remaining factor is bounded by
$\eta^{-1}$, so the conditional variance tends to zero as
$N\to\infty$. This establishes the required concentration
for the sequential limit used in the main text. It does not,
by itself, establish a local law down to the mean level
spacing. Such a statement requires additional stability
estimates for the Dyson equation. In particular, boundedness
of the smoothed density cannot be assumed uniformly in
$\eta$ near a divergent central density.
The independence used in this calculation belongs to the
Gaussian approximation to the bath fluctuations. The exact
Wishart entries are not independent, their higher cumulants
give corrections beyond the leading large-$m$ theory.

The mean \eqref{eq:Qmean} still contains the resolvent $G^{(j)}$ of the cavity system rather than
that of the full one. However, removing a single level changes the diagonal elements only weakly.
Indeed, the same block inversion that gives \eqref{eq:cavity} also yields, for $k,l\ne j$, the
exact identity $G_{kl}=G^{(j)}_{kl}+G_{kj}G_{jl}/G_{jj}$. Combining it with the Ward identity, we
obtain
\begin{equation}\label{eq:cavdiff}
\Big|\frac1N\sum_{k\ne j}S(E_j,E_k)\big(G_{kk}-G^{(j)}_{kk}\big)\Big|
\le\frac{2}{mN}\,\frac{\sum_k|G_{kj}|^2}{|G_{jj}|}
=\frac{2}{mN}\,\frac{|\mathrm{Im}\,G_{jj}|}{\eta\,|G_{jj}|}\le\frac{2}{mN\eta}.
\end{equation}
Hence the sum in \eqref{eq:Qmean} changes only by $O\big(1/(N\eta)\big)$ when $G^{(j)}_{kk}$ is
replaced by $G_{kk}$ and the term $k=j$ is restored. The concentration over $y_j$ alone
does not yet fix $G_{jj}$, because the weighted trace $\frac1N\sum_kS(E_j,E_k)G_{kk}$ is itself random.
At fixed $\eta>0$ it concentrates about its mean as $N\to\infty$, as for any Wigner-type matrix with a
bounded variance profile~\cite{Ajanki2017}. Together with the estimates above this gives the
concentration of $G_{jj}$ about $g_j$, and averaging yields the closed system
\eqref{eq:gj}.

\section{The edge of the distribution: derivations}\label{app:edge}
In this appendix we derive the results of Sec.~\ref{sec:edge}. The notation is that of
Eq.~\eqref{eq:Top}.

\subsection{The square-root law}\label{app:sqrt}
For a flat profile, $S=2/m$, Eq.~\eqref{eq:Top} reduces to the quadratic equation \eqref{eq:scalar}
for a single number $\avE{g}$. For real $b$ outside the support it has two real roots, the physical
one, $\avE{g}\simeq1/b$ at large $b$, and an unphysical one. At $b_e$ the discriminant vanishes and the
two roots merge. Inside the support they form a complex-conjugate pair, and
$\mathrm{Im}\,\avE{g}\propto\sqrt\delta$. 

Let us write $g=g_r-ig_i$, so that $\rst=\avE{g_i}/\pi$. At $\lambda=b$, Eq.~\eqref{eq:Top} splits into
\begin{equation}\label{eq:realpair}
\frac{g_r}{g_r^2+g_i^2}=b-\TT[g_r],\qquad\qquad
g_i=\big(g_r^2+g_i^2\big)\,\TT[g_i].
\end{equation}
Outside the support $g_i=0$, and the first equation alone determines $g_r$. Near the edge $g_i$ is
small, and we write it as $g_i=\vartheta\,w(E)$, with a small amplitude $\vartheta$ and a shape $w(E)$
that stays finite at the edge. The normalisation of $w$ is arbitrary, and the result \eqref{eq:root}
below does not depend on it. To first order in $\vartheta$ the second equation is linear, and $g_r$
may be taken at the edge, which gives the eigenvalue problem \eqref{eq:thresh}. A non-zero density can appear only when the operator $g_r^2\TT$ has
the eigenvalue one, and the density on each level is then sustained by the density on the others.
Far from the band $g_r\simeq1/b$ is small, and so are all eigenvalues of $g_r^2\TT$. As $b$
decreases, $g_r$ grows, and the edge $b_e$ is the point at which the largest eigenvalue reaches one.
The kernel of $g_r^2\TT$, $g_r(E)^2\nu(E')S(E,E')$, is positive, so by the Perron--Frobenius
(Jentzsch) theorem its largest eigenvalue is simple and has a positive eigenfunction, $w>0$, as it
must be for the shape of $g_i=-\mathrm{Im}\,g\ge0$. For a flat
profile $w=1$, $g_e=\sqrt{m/2}$ and $b_e=2\sqrt{2/m}$.

The linear equation fixes the shape $w$ but not the amplitude $\vartheta$. The relation between
$\vartheta$ and $\delta$ follows from the next order. Since $g_i$ enters \eqref{eq:realpair} only
through $g_i^2$, the corrections to $g_r$ and to $b$ are of order $\vartheta^2$,
\begin{equation}
g_r=g_e+\vartheta^2g_1(E),\qquad \delta=b_e-b=\vartheta^2d .
\end{equation}
Expanding the left-hand side of the first equation, $g_r/(g_r^2+g_i^2)\simeq1/g_r-g_i^2/g_r^3$, we
obtain at order $\vartheta^2$
\begin{equation}\label{eq:v2}
\frac{g_1}{g_e^2}-\TT[g_1]=d-\frac{w^2}{g_e^3} .
\end{equation}
The operator on the left, $\mathcal K[f]=f/g_e^2-\TT[f]$, is symmetric like $\TT$ and annihilates $w$ by
\eqref{eq:thresh}. Multiplying \eqref{eq:v2} by $w$ and averaging over the band therefore eliminates
the unknown $g_1$, $\avE{w\,\mathcal K[g_1]}=\avE{g_1\,\mathcal K[w]}=0$, and leaves a condition on the right-hand side
alone, $\avE{w\,(d-w^2/g_e^3)}=0$, which gives
\begin{equation}
d=\frac{\avE{w^3/g_e^3}}{\avE{w}} .
\end{equation}
Hence $\vartheta=\sqrt{\delta/d}$, and the density $\rst=\avE{g_i}/\pi=\vartheta\avE{w}/\pi$ becomes Eq.~\eqref{eq:root}.
Both averages are positive, so $d>0$ and $A>0$, and $\vartheta$ is real inside the support,
$\delta>0$.

\subsection{\texorpdfstring{$s\gg W$}{s >> W}}\label{app:lsedge}
At $s\gg W$ the eigenvalue problem \eqref{eq:thresh} is not needed: the solution of \eqref{eq:Top}
stays close to the scalar one of Sec.~\ref{sec:larges}, and the edge can be read off directly. By
\eqref{eq:Slarge} the profile is flat up to terms of order $s^{-2}$, so $g(E,\lambda)$ differs from
its band average $\avE{g}$ by terms of the same order. The constant part $2/m$ of $S$
does not act on the difference $g-\avE{g}$, whose band average vanishes, and, exactly,
\begin{equation}\label{eq:lsT}
\TT[g](E)=u(E)\,\avE{g}+\int\dd E'\,\nu(E')\Big[S(E,E')-\frac2m\Big]\big[g(E')-\avE{g}\big]
\simeq u(E)\,\avE{g} .
\end{equation}
The last term is a product of two factors of order $s^{-2}$, and we omit it. A level at energy $E$
then feels the band only through $\avE{g}$, with the coupling $u(E)$ of the sum rule
\eqref{eq:sumrule}. This is the structure of Sec.~\ref{sec:smalls}, with $\avE{g}$ in place of the
local $g(E,\lambda)$. Averaging $g=1/(\lambda-u\,\avE{g})$ over the band closes the equation for
$\avE{g}$. The coupling varies across the band only by terms of order $s^{-2}$, and we expand around
its mean $\avE{u}$:
\begin{equation}\label{eq:lsg}
\avE{g}=\avE{\frac{1}{\lambda-u\,\avE{g}}}\simeq\frac{1}{\lambda-\avE{u}\,\avE{g}} .
\end{equation}
The term linear in $u-\avE{u}$ vanishes on averaging, and the error is of second order in
$u-\avE{u}$, i.e.\ of order $s^{-4}$. The expansion holds up to the edge itself, because the
denominator $\lambda-\avE{u}\,\avE{g}\simeq1/\avE{g}$ never vanishes: the square root of $\rst$ comes
from the solution $\avE{g}$, not from the denominators.

Equation~\eqref{eq:lsg} is the scalar equation \eqref{eq:scalar} with $2/m\to\avE{u}$. Its solution
is the semicircle \eqref{eq:larges} of variance $\avE{u}$, which is the exact width,
$\avE{u}=\sigma_\beta^2$, Eq.~\eqref{eq:sumrule}. The edge and the amplitude are therefore those of
this semicircle, up to relative corrections of order $s^{-4}$. By \eqref{eq:Slarge},
$\avE{u}\simeq(2/m)\big[1-\avE{(E-E')^2}/s^2\big]=(2/m)(1-2\sigma_\nu^2/s^2)$, where
$\sigma_\nu^2=\avE{E^2}-\avE{E}^2$ is the variance of the levels distributed with the density $\nu$.
With $b_e=2\sigma_\beta$ and $A=1/(\pi\sigma_\beta^{3/2})$ this gives Eqs.~\eqref{eq:lsprofiles} and
\eqref{eq:lsA}. For the GUE $\sigma_\nu^2=1$, and $\avE{u}$ agrees with $\Psi\simeq1-2/s^2$, Eq.~\eqref{eq:asym}.

\subsection{\texorpdfstring{$s\ll W$}{s << W}: the layer equation}\label{app:layer}
Here $H_0\in$ GUE, $\nu=\rsc$, and the notation is that of Sec.~\ref{sec:layer}.
Multiplying $1/g=b-\TT[g]$ by $g$ and
using $\TT[1]=u$, Eq.~\eqref{eq:sumrule}, we write the vector Dyson equation \eqref{eq:Top} exactly as
\begin{equation}\label{eq:layer}
u(E)\,g^2-b\,g+1=g(E)\int\dd E'\,\nu(E')\,S(E,E')\,\big[g(E)-g(E')\big].
\end{equation}
The left-hand side is the local scalar equation \eqref{eq:localg}, here with the exact coupling
$u(E)$. Its double root at $b=2\sqrt{u(E)}$ is the edge of the local semicircle. The right-hand side
couples the levels and vanishes when $g$ does not depend on $E$. Near the centre
$u(E)\simeq u_0\,(1-E^2/8)$, with $u_0=u(0)\simeq(2s/m)(1-s/2)$ from \eqref{eq:u}. At $b=2\sqrt{u_0}$ the
central level sits exactly at its own edge, with the double root $g=1/\sqrt{u_0}$. We measure $g$ and
$b$ from this point,
\begin{equation}\label{eq:layervars}
g=\frac{1+\varphi}{\sqrt{u_0}},\qquad b=2\sqrt{u_0}\,(1-s\,q),
\end{equation}
so that $s\,q$ is the relative distance from the edge of the central level. Inserting
\eqref{eq:layervars} in the left-hand side and keeping terms of second order in the small quantities
$\varphi$, $E$ and $s\,q$,
\begin{equation}
\Big(1-\frac{E^2}8\Big)(1+\varphi)^2-2(1-s\,q)(1+\varphi)+1\simeq\varphi^2-\frac{E^2}{8}+2s\,q .
\end{equation}
For the right-hand side we need the action of the Lorentzian on a function $f$ that varies slowly on
the scale $s$,
\begin{equation}\label{eq:absd}
\frac1\pi\int\dd E'\,\frac{s\,[f(E)-f(E')]}{(E-E')^2+s^2}\simeq s\,|\partial_E|f(E),\qquad
|\partial_E|f(E)=\frac1\pi\dashint\dd E'\,\frac{f(E)-f(E')}{(E-E')^2},
\end{equation}
where $|\partial_E|$ is the operator with the Fourier symbol $|k|$. With $\nu(E')\simeq\nu(0)$ the
right-hand side of \eqref{eq:layer} is $g\,(2\pi s\,\nu(0)/m)\,s\,|\partial_E|g$. Its prefactor
$2\pi s\,\nu(0)/m$ is the local coupling $u_0$ itself, since the Lorentzian of area $\pi s$ collects the
neighbours with the density $\nu(0)$. The density therefore cancels, and it remains only in the
curvature of $u(E)\propto\nu(E)$, the term $E^2/8$. Altogether, Eq.~\eqref{eq:layer} near the centre
reads
\begin{equation}\label{eq:layerexp}
\varphi^2-s\,|\partial_E|\varphi=\frac{E^2}{8}-2s\,q .
\end{equation}
The term $\varphi^2$ makes two roots merge, as in the scalar equation, $E^2/8$ is the decrease of the
local radius away from the centre, and $q$ is the distance from the edge. The coupling
$s\,|\partial_E|\varphi$ is of first order in $s$. Dropping it decouples the levels, and
Eq.~\eqref{eq:layerexp} is solved level by level,
\begin{equation}\label{eq:outer}
\varphi_{\rm out}(E)=-\Big(\frac{E^2}{8}-2s\,q\Big)^{1/2},
\end{equation}
with the sign of the physical root, $g\simeq1/b$ far from the band, and $\mathrm{Im}\,g<0$ where the
root is complex. We call it the outer solution. For $q<0$ all levels are outside their local spectra,
and $\varphi_{\rm out}$ is real. For $q>0$ the levels with $E^2<16s\,q$ are inside them, and
$-\mathrm{Im}\,g=(2s\,q-E^2/8)^{1/2}/\sqrt{u_0}$. The density is then
\begin{equation}\label{eq:outerrho}
\rst=\frac{\nu(0)}{\pi\sqrt{u_0}}\int\dd E\,\Big[2s\,q-\frac{E^2}8\Big]_+^{1/2}
=\frac{2\sqrt2\,\nu(0)}{\sqrt{u_0}}\,s\,q=\frac{\sqrt2}{\pi u_0}\,\big(2\sqrt{u_0}-b\big) .
\end{equation}
This is the linear edge of the $s\to0$ law, with the slope $4\sqrt2/(\pi R_0^2)$ of
Eq.~\eqref{eq:localsc} at $b\to R_0$, since $R_0^2\simeq4u_0$. The outer solution thus reproduces the picture of independent local
semicircles.

The outer solution is self-consistent only if the
term we dropped is small on it. At $q=0$ it has a corner at the centre, $\varphi_{\rm out}=-|E|/\sqrt8$.
For the corner $|\partial_E||E|\simeq-(2/\pi)\ln(1/|E|)$ at $|E|\ll1$, where the band cuts the integral
off at $|E'|\sim1$. Hence
\begin{equation}\label{eq:neglected}
-s\,|\partial_E|\varphi_{\rm out}\simeq-\frac{s}{\pi\sqrt2}\,\ln\frac1{|E|} .
\end{equation}
The neglected term grows towards the centre, while the retained term $E^2/8$ decreases. The two are
comparable at $E^2\sim s$, up to the logarithm. Near the centre, precisely where the edge is formed,
the coupling cannot be neglected, however small $s$ is. At the boundary of this region,
$|E|\sim\sqrt s$, the solution is still close to the outer one, $\varphi\simeq-|E|/\sqrt8\sim\sqrt s$, so
inside it $\varphi\sim\sqrt s$ as well. All terms of \eqref{eq:layerexp} are then of the same order,
$\varphi^2\sim E^2\sim s$ and $s\,|\partial_E|\varphi\sim s\,\varphi/\sqrt s\sim s$, and the term $2s\,q$
joins them for $q\sim1$. The edge region is therefore $q\sim1$, i.e.\ $\delta\sim sR_0\sim s\,\sigma_\beta$.

This is the situation of a boundary layer in hydrodynamics. A small viscosity is negligible in the
bulk of a flow, but not in a thin layer, of thickness proportional to the square root of the
viscosity, where the inviscid solution fails. Here the small parameter is $s$, the outer solution is
that of independent local semicircles, and the layer has the width $\sqrt s$. Such problems are solved
by matched asymptotic expansions~\cite{BenderOrszag1999}. Inside the layer the variables are rescaled
so that the small term is as large as the others, and the inner solution is matched to the outer
one. The analogy is not complete in two respects. Our layer lies inside the band, not at its
boundary, like the viscous layer of a shock in Burgers' equation, and the term that smooths it is
nonlocal.

Inside the layer we use the stretched variables $E=\sqrt s\,y$ and
$\varphi=\sqrt s\,\psi$. Since $|\partial_E|=|\partial_y|/\sqrt s$, every term of \eqref{eq:layerexp} is
proportional to $s$, and after division by $s$
\begin{equation}\label{eq:inner}
\psi^2-|\partial_y|\psi=\frac{y^2}{8}-2q .
\end{equation}
The coupling is now of the same order as the other terms, and $s$ has disappeared, as the viscosity
disappears from the rescaled equations of a viscous boundary layer. The inner solution must match the outer
one far from the centre, $\psi\simeq-|y|/\sqrt8$ for $|y|\to\infty$ at fixed $q$.

Equation~\eqref{eq:inner} is not yet complete, because the nonlocal operator is not regular on this
tail. At large $|y'|$ the integrand of $|\partial_y|\psi$ is $[\psi(y)-\psi(y')]/(y-y')^2\simeq|y'|/(\sqrt8\,y'^2)$,
and both sides of $y'=0$ give $(2/\pi\sqrt8)\int\dd y'/y'$, which diverges logarithmically. This is the
logarithm \eqref{eq:neglected}: the stretched variables have pushed the distant levels to
$|y'|\to\infty$, where the band, not the layer, cuts the integral off. The layer equation must
therefore be completed by the contribution of the distant levels, taken from the outer solution.

We split the coupling integral in \eqref{eq:layer} at $|E'|=\Lambda$,
in the overlap region $\sqrt s\ll\Lambda\ll1$, where both descriptions hold and $\varphi\simeq-|E|/\sqrt8$.
For a level in the layer, $|E|\ll\Lambda$, we write the right-hand side of \eqref{eq:layer} as
$I_<+I_>$, the contributions of $|E'|<\Lambda$ and $|E'|>\Lambda$:
\begin{equation}\label{eq:IinIout}
\begin{aligned}
I_<&=g(E)\int_{|E'|<\Lambda}\dd E'\,\nu(E')\,S(E,E')\,\big[g(E)-g(E')\big]\\
&\simeq\frac s\pi\dashint_{|E'|<\Lambda}\dd E'\,\frac{\varphi(E)-\varphi(E')}{(E-E')^2}
=s\,|\partial_y|_\Lambda\psi(y),\\
I_>&=g(E)\int_{|E'|>\Lambda}\dd E'\,\nu(E')\,S(E,E')\,\big[g(E)-g(E')\big]\\
&\simeq\frac{2s^2\nu(0)}{m\,u_0}\int_{|E'|>\Lambda}\frac{\dd E'}{E'^2}\,\frac{\nu(E')}{\nu(0)}
\Big[1-\sqrt{u_0}\,g_{\rm loc}(E')\Big]=s\,C(\Lambda).
\end{aligned}
\end{equation}
In the inner region we used, as above, $\nu(E')\simeq\nu(0)$, the form \eqref{eq:absd} of the
Lorentzian and $g=(1+\varphi)/\sqrt{u_0}$, and then the stretched variables $E'=\sqrt s\,y'$,
$\varphi=\sqrt s\,\psi$. This gives the truncated operator
\begin{equation}\label{eq:dtrunc}
|\partial_y|_\Lambda\psi(y)=\frac1\pi\dashint_{-y_\Lambda}^{y_\Lambda}\dd y'\,\frac{\psi(y)-\psi(y')}{(y-y')^2},\qquad
y_\Lambda=\frac{\Lambda}{\sqrt s}.
\end{equation}
In the outer region $|E'|\gg|E|$, so that $S\simeq(2/m)\,s^2/E'^2$, $g(E)\simeq1/\sqrt{u_0}$, and
$g(E')\simeq g_{\rm loc}(E')$ is the local root over the whole semicircle. With $2\pi s\,\nu(0)/m=u_0$ the
prefactor is $s/\pi$, which gives the constant $s\,C(\Lambda)$ below. The integrand is even in $E'$, and
the two sides $E'\gtrless0$ give the factor $2$ in its second form.

The truncated operator grows logarithmically with its limit. Its finite part defines the
regularised operator,
\begin{equation}\label{eq:dreg}
|\partial_y|_{\rm reg}\,\psi(y)=\lim_{y_\Lambda\to\infty}\Big[|\partial_y|_\Lambda\psi(y)-\frac{\ln y_\Lambda}{\pi\sqrt2}\Big],
\qquad
|\partial_y|_\Lambda\psi\simeq|\partial_y|_{\rm reg}\,\psi+\frac{1}{\pi\sqrt2}\ln\frac{\Lambda}{\sqrt s}.
\end{equation}
The limit exists because the subtracted logarithm is exactly the contribution of the tail
$\psi\simeq-|y|/\sqrt8$, and it does not depend on $y$ for $|y|\ll y_\Lambda$. The outer constant is
\begin{equation}\label{eq:Cout}
C(\Lambda)=\frac1\pi\int_{|E|>\Lambda}\frac{\dd E}{E^2}\,\frac{\nu(E)}{\nu(0)}\Big[1-\sqrt{u_0}\,g_{\rm loc}(E)\Big]
=\frac2\pi\int_\Lambda^2\frac{\dd E}{E^2}\,\bigg[\frac{\nu(E)}{\nu(0)}-1+\Big(1-\frac{\nu(E)}{\nu(0)}\Big)^{1/2}\bigg].
\end{equation}
Here $\nu(E)/\nu(0)=u(E)/u_0=(1-E^2/4)^{1/2}$, and the second form follows from the local root
$g_{\rm loc}=[b-(b^2-4u)^{1/2}]/2u$ at $b=2\sqrt{u_0}$. Near the centre the square bracket is
$\simeq E/\sqrt8$, the same linear tail, so $C(\Lambda)$ grows logarithmically as $\Lambda\to0$. Separating the logarithm,
\begin{equation}\label{eq:Coutval}
\begin{gathered}
C(\Lambda)=\frac{1}{\pi\sqrt2}\ln\frac1\Lambda+C_{\rm out},\\
C_{\rm out}=\frac2\pi\int_0^2\frac{\dd E}{E^2}\bigg[\frac{\nu(E)}{\nu(0)}-1+\Big(1-\frac{\nu(E)}{\nu(0)}\Big)^{1/2}-\frac{E}{\sqrt8}\,\theta(1-E)\bigg]
=-\frac12+\frac{1+3\ln2-\ln(1+\sqrt2)}{\pi\sqrt2},
\end{gathered}
\end{equation}
where the integral is elementary with $E=2\sin t$.

Replacing the coupling $s\,|\partial_E|\varphi$ in \eqref{eq:layerexp} by $I_<+I_>$ from
\eqref{eq:IinIout}, using \eqref{eq:dreg} and \eqref{eq:Coutval}, and dividing by $s$,
\begin{equation}
\psi^2-|\partial_y|_{\rm reg}\,\psi=\frac{y^2}{8}-2q+\frac{1}{\pi\sqrt2}\Big[\ln\frac{\Lambda}{\sqrt s}+\ln\frac1\Lambda\Big]+C_{\rm out}
=\frac{y^2}{8}-2q+\frac{\ln(1/s)}{2\pi\sqrt2}+C_{\rm out}.
\end{equation}
The logarithm that the inner part loses at $|E'|=\Lambda$ is exactly the one that the outer part gains
below $\Lambda$, since both come from the same linear tail in the overlap region. The arbitrary
division point drops out. What remains is the logarithm \eqref{eq:neglected} evaluated at the width
of the layer, $|E|\sim\sqrt s$. Absorbing the constants into the distance from the edge gives the layer
equation \eqref{eq:universal}. At given $Q$ it contains neither $s$ nor a cutoff, and $s$ enters only through the shift of $Q$ against $q$. The regularisation of $|\partial_y|$ is thus not an ad hoc
subtraction: the logarithm removed from the operator reappears as the shift of $Q$ against $q$. The
outer levels push the edge inwards by a logarithmically enhanced amount, and the layer adds a
correction of order one in $Q$.

Equation~\eqref{eq:universal} is a stationary Benjamin--Ono equation
with a quadratic potential. The equation is nonlinear and nonlocal. We know of no closed-form solution
and solve it numerically. The parameter $Q$ plays the role of the distance from the edge: it grows as $b$ moves into
the spectrum. Outside the spectrum the solution $\psi(y)$ is real and $\rst=0$, so the edge $Q_*$ is the
largest $Q$ at which a real solution exists. Without the coupling term the edge is at $Q=0$: the local
root $\psi=-(y^2/8-2Q)^{1/2}$ is real for all $y$ as long as $Q\le0$. For $Q>0$ it is complex in the
window $y^2<16Q$, which opens at the centre as the central level enters its own spectrum. If the coupling is
dropped altogether, including the distant levels that shift $Q$ against $q$ in \eqref{eq:universal},
then $Q=q$, and since $s\,q\simeq(R_0-b)/R_0$ this is the window $E^2<16(R_0-b)/R_0$ of independent
levels found at the beginning of this section. The coupling moves the edge to
$Q_*>0$. There the real solution ceases to exist in the same way as in the scalar equation: two real
solutions merge, and beyond this point they become complex. At the merging point the equation
linearised about $\psi$, $2\psi w=|\partial_y|w$, has a nonzero solution $w(y)>0$. This is the
threshold condition \eqref{eq:thresh} of Sec.~\ref{sec:sqrt} in the variables of the layer, and $w$ is the
edge mode of Sec.~\ref{sec:sqrt}, now as a function of $y$. Just beyond the edge
$\mathrm{Im}\,\psi\propto\sqrt{Q-Q_*}\,w$, the square root of Sec.~\ref{sec:sqrt}. Numerically we follow
the real solution as $Q$ grows and locate $Q_*$ as the point where the linearised operator acquires a
zero eigenvalue. Its eigenvector is $w$. Returning to \eqref{eq:layervars},
\begin{equation}\label{eq:layeredge}
b_e=2\sqrt{u_0}\,\Big[1-s\Big(Q_*+\frac{\ln(1/s)}{4\sqrt2\,\pi}+\frac{C_{\rm out}}2\Big)\Big],\qquad
\delta\simeq sR_0\,(Q-Q_*).
\end{equation}
The edge mode is concentrated in the layer. Inserting it and $g_e\simeq2/R_0$ in \eqref{eq:root}, with
$\rsc\simeq1/\pi$ across the layer, gives
\begin{equation}\label{eq:layerA}
A=\frac{2^{3/2}}{\pi^2}\,\frac{\big(\int w\,\dd y\big)^{3/2}}{\big(\int w^3\,\dd y\big)^{1/2}}\,\frac{\sqrt s}{R_0^{3/2}}
=a_0\,\frac{m^{3/4}}{s^{1/4}},\qquad
a_0=\frac{\big(\int w\,\dd y\big)^{3/2}}{2^{3/4}\pi^2\big(\int w^3\,\dd y\big)^{1/2}}.
\end{equation}
The factor $\sqrt s$ is the fraction of levels in the layer. The numerical solution gives
$Q_*\approx0.0276$ and $a_0\approx0.254$.
With $2\sqrt{u_0}\simeq R_0\,(1-s/4)$ Eqs.~\eqref{eq:layeredge} and \eqref{eq:layerA} give
Eq.~\eqref{eq:small}.

\end{appendix}

\bibliography{references}

\end{document}